# Unpacking Human-AI Interaction in Safety-Critical Industries: A Systematic Literature Review

**Tita A. Bach[1], Jenny K. Kristiansen[1], Aleksandar Babic[1], and Alon Jacovi[2]**
[1]Group Research and Development, DNV, Høvik, Norway
[2]Bar Ilan University, Ramat Gan, Israel

Corresponding author: Tita A. Bach (e-mail: tita.alissa.bach@dnv.com).

**ABSTRACT** Ensuring quality human-AI interaction (HAII) in safety-critical industries is essential. Failure to do so can lead to catastrophic and deadly consequences. Despite this urgency, existing research on HAII is limited, fragmented, and inconsistent. We present here a survey of that literature and recommendations for research best practices that should improve the field. We divided our investigation into the following areas: 1) terms used to describe HAII, 2) primary roles of AI-enabled systems, 3) factors that influence HAII, and 4) how HAII is measured. Additionally, we described the capabilities and maturity of the AI-enabled systems used in safety-critical industries discussed in these articles. We found that no single term is used across the literature to describe HAII and some terms have multiple meanings. According to our literature, seven factors influence HAII: user characteristics (e.g., user personality), user perceptions and attitudes (e.g., user biases), user expectations and experience (e.g., mismatched user expectations and experience), AI interface and features (e.g., interactive design), AI output (e.g., perceived accuracy), explainability and interpretability (e.g., level of detail, user understanding), and usage of AI (e.g., heterogeneity of environments). HAII is most measured with user-related subjective metrics (e.g., user perceptions, trust, and attitudes), and AI-assisted decision-making is the most common primary role of AI-enabled systems. Based on this review, we conclude that there are substantial research gaps in HAII. Researchers and developers need to codify HAII terminology, involve users throughout the AI lifecycle (especially during development), and tailor HAII in safety-critical industries to the users and environments.

## I. INTRODUCTION

Artificial Intelligence (AI) has rapidly become widely used across all industries in recent years [1]. This is especially true in safety-critical industries [2], [3]. Poor human-AI interaction (HAII) is one of the major reasons AI is underused, misused, or overused, resulting in catastrophic outcomes [4], [5], [6], [7], [8], [9]. Ensuring high quality HAII is thus an essential prerequisite to successfully implementing and adopting AI in these areas.

In this review, we examine the HAII literature with a focus on the influence of HAII in safety-critical industries. We define safety-critical industries as "industries in which safety is of paramount importance and where the consequences of failure or malfunction may be loss of life or serious injury, serious environmental damage, or harm to plant or property [10], [11]." These industries may include, for example, construction [12], healthcare [13], energy [14], aviation [15], space [16], and transportation [17]. Because of the safety-critical nature, AI implementation in these industries requires special consideration.

In attempting to review best practices for HAII, we quickly found that HAII terminology is not yet consistent across the research. Therefore, in this review we refer to HAII as any situation where a human is directly using, engaging, or interacting with AI to complete a task or achieve a goal (the question of defining AI is discussed in the next section).

A common definition of HAII is important because the quality of HAII can determine whether users (i.e., the humans involved) are able to realize the purpose of the AI for their benefit and for the benefit of society at large [18], [19]. Effective HAII can also encourage users to continue using the AI and form human-AI relationships. It is difficult to understand and research HAII without a clear definition.

HAII in safety-critical industries, as opposed to HAII generally, has several properties that present unique challenges [3], [10], [11], [20], yet the literature in this area is fragmented across many isolated sub-areas. This diversity is partly a result of AI being used for many different applications, users, and environments in safety-critical industries. Developers and researchers therefore need to tailor their AI to users and their environments and involve those users throughout the development process. To support that need, we categorized the primary roles of AI in this review and investigated the most common methods for measuring HAII.

To our knowledge, this review is the first study of AI which aims to explore what HAII is, what factors play a role in HAII, and what gaps exist in the literature that is motivated by understanding how AI will impact safety-critical industries. Our goal is to identify the commonalities across seemingly separate lines of HAII research so that AI developers and researchers across industries can benefit and learn from each other. In addition, with this review we present methods for fostering more conducive HAII in safety-critical settings and recommend best practices for future research.

We investigated the literature on HAII in safety-critical industries by asking the following research questions (RQ):

1) RQ1: What term is used to describe the interaction between humans and an AI-enabled system?
2) RQ2: What is the primary role of the AI-enabled system in the interaction?
3) RQ3: What factors were found to influence HAII?
4) RQ4: How is HAII being measured in safety-critical industries?

We used a database search strategy to answer these questions. We first developed inclusion criteria appropriate for studying HAII in safety critical industries. We then conducted a targeted search of four scientific databases and screened the resulting articles against our inclusion criteria to ensure that they focused on the interaction between users and an AI-enabled system, applied to a safety-critical industry, and involved users who intended to use the AI-enabled system.

We also investigated the capabilities and maturity of the AI-enabled systems found in the included articles. These determine which tasks the system can perform, and thus affect the nature of HAII. The maturity level refers to where in the implementation phase the system is and describes the environment in which the HAII happens and who will be using it, which influences the quality and dynamic of HAII.

## II. BACKGROUND

### A. WHAT IS AI? Definition and risks of AI-enabled systems

This section is divided into discussions about 1) AI as a term with inconsistent use and definitions, and our chosen term "AI-enabled system" and its description, and 2) the risks of AI-enabled systems. Establishing a consistent definition for AI in this review allows us to discuss this topic which spans many disparate fields. Clarifying the term and its description also shows what this review does and does not cover.

#### 1) INCONSISTENT USE OF THE TERM

"AI" has been used and defined in various ways across fields and literature. This is concerning because without a common definition of what AI is, researchers from different fields, or even within the same field, can end up talking past each other without contributing to our greater understanding. Another concern is that the broad range of AI definitions, especially in official documents, makes it difficult to determine which products are considered AI and therefore which regulations apply to them [21], [22], [23]. For example, [24] found that there is great diversity across policy documents in the use of terminologies such as "AI," "ML," "Deep Learning," and other similar terms.

The British Standards Institution (BSI) highlighted a related concern in their introduction to the European Artificial Intelligence Draft Act (AIA) titled "Ethical and trustworthy artificial intelligence [25]." That report presented eight different AI definitions in documents published by international and national regulatory bodies including WHO, OECD, US House of Representatives, Gov.UK, The National Information Security Standardization Technical Committee of China, and the EU AI Act [22], [25]. Adding further ambiguity to the definition of AI, they found that definitions of AI usually include terms such as "intelligence," "automation," "autonomy," "authority," "perceiving," and others which can be interpreted in various ways [22], [25].

Making things even worse, in many cases these terms are used implicitly without providing any underlying explicit definitions. AI is difficult to define [26], but defining and describing AI can not only benefit policymakers and regulators whose interest lie in governing and regulating AI, but also supports AI research and communities coordinating their collective efforts to move the field forward.

To facilitate discussion of AI in this review, we use the term "AI-enabled system" to denote "any system that contains or relies on one or more AI components [27]." In this case, an AI component refers to a distinct unit of software that performs a specific function or task within an AI-enabled system and consists of a set of AI elements (i.e., models, data, algorithms), which, through implementation, creates an AI component. This term and its description are taken from DNV-RP-0671 "Assurance of AI-enabled systems [27]," a Recommended Practice (RP) being developed by DNV, an independent, third-party expert in assurance and risk management with a focus on safety-critical industries. DNV and the RP share the same goal of building AI trustworthiness as the proposed AI Act [28]. The RP is being developed for application in safety-critical industries by subject-matter experts. At the time of writing this review, the RP is expected to be published in Autumn 2023. We consider this description of AI-enabled systems to be both comprehensive and generalizable to all relevant AI-enabled systems, but also specific and concrete enough to form an effective definition of AI. Importantly, this description is aligned with our research questions.

#### 2) RISKS OF AI-ENABLED SYSTEMS

The first step of developing an AI-enabled system that is responsible and ethical is justifying why the AI-enabled system is needed in the first place [4], [23]. Often these systems carry inherent technical risks such as difficulty predicting its behaviors and tracking errors in its processes (i.e., the "black box" problem) [2], [3], [29], [30]. In addition, output from AI-enabled systems are rarely 100% accurate and the accuracy often changes throughout each life-cycle [30]. Growing research into the biases of AI-enabled systems has also highlighted the risks of discrimination [31], [32], [33].

Implementing an AI-enabled system in safety-critical industries carries additional risks and challenges due to the level of impact on human lives [2], [3], [27]. Therefore, an essential first step is justifying that the benefits of implementing an AI-enabled system outweigh the risks.

An important step after justifying the use of an AI-enabled system is to work closely with users throughout the development lifecycle. This will help developers understand the function and limitations of their AI-enabled systems and can minimize the risks of these biases by, for example, training the algorithm on a diverse range of users and situations [34].

### *B. WHO ARE THE HUMANS? Users and their humane characteristics*

The humans in HAII are defined as the target end-users ("users") who use, engage with, or interact *directly* with an AI-enabled system. For example, a radiologist who gets support from an AI-enabled system identifying tumors in MRI images or a pilot who uses an AI-enabled system to find ideal navigation routes. Users are the main stakeholders and an integral part of ensuring that AI-enabled systems are used, and their benefits realized.

Including users in the development phase of AI-enabled systems is crucial, especially because developers and users can have different priorities for AI-enabled systems [35]. In practice this is rarely done. Instead, development usually happens in isolation away from the users and even UX designers [36]. This a growing problem as AI-enabled systems become more and more embedded in daily life and users may not even realize that they are using them [37]. Ethical HAII should not only inform users that they are interacting with AI-enabled systems but also ensure they understand the AI-enabled system presented to them as the developers intended [18], [28]. Transparency and informed consent, as well as users' AI literacy, are thus important topics [28], [38].

Definitively identifying the users of an AI-enabled system during the development phase can be a challenge. Currently, AI-enabled systems are typically specialized to solve one problem or perform one task, but its developers hope it will be able to be generalized across a broad spectrum of users. This makes it difficult to identify those users. Local implementation will most likely need significant adjustments (e.g., technical and organizational adjustments) to match the profile of users to specific operational environments [23].

For example, effectively integrating an AI-enabled system for assisting radiologists in classifying breast cancer into the IT infrastructure of the local healthcare organization and adjusting it to fit into the existing clinical workflow and the patient population is a good way to foster quality HAII. As part of this process, the local healthcare organization must first determine if the AI model is appropriate for their radiologists, which may have a different profile than those used to develop the AI-enabled system.

Although user modeling can be employed to simulate the user profiles best suited to the AI-enabled system in question [38], there is always a possibility of discrepancies between the modeled user profile and the real-world user profile (i.e., actual users who use the AI-enabled system in question in the real-world) [23]. To avoid unanticipated HAII, we need a deep understanding of the profiles of the users and the operational environment where the AI-enabled system will be implemented [38], [39]. Knowledge of both will allow us to understand the context and use case to improve the quality of HAII.

Importantly, understanding users of AI-enabled systems requires us to consider the unique human characteristics that influence how people interact with AI-enabled systems, such as human judgement and decision-making [40], [41]. If we want to make decisions with AI-enabled systems, we need to understand how people think, otherwise users will make incorrect decisions which can have catastrophic consequences in these safety-critical industries.

One particular aspect of human thinking that strongly influences HAII is our ability to apply heuristic strategies to quickly form judgements in order to reduce cognitive effort, also called *thinking fast* [42]. This is a result of a biological limit in processing knowledge and information [43], [44]. In cases when a decision needs to be made quickly and familiarity is an asset, for example, pilots quickly assessing AI recommendations on navigation routes and doctors cross-referencing AI suggestions to make a diagnosis, a heuristic strategy is effective and efficient. In such cases, thinking fast can be very beneficial for experienced persons in a relatively simple scenario in a highly familiar area [42].

In other cases, however, this strategy can lead to biases in decision-making such as availability bias [45]. This is when we overestimate a topic, event, or something as more important than others just because it comes to mind easily or the solution is readily available. For example, a recent minor incident may draw the attention of aircraft maintenance resources to prevent it from happening again while other less frequent but more impactful problems are ignored. In addition, our limited capacity to process information may also lead to an overreliance on AI-enabled systems which leads users to blindly accept and follow AI suggestions, even if the suggestions are wrong, rather than listening to own or alternative suggestions [46], [47].

Mechanisms of AI overreliance can come in the form of, for example, automation bias (i.e., favoring recommendations from automated systems, such as AI-enabled systems, and ignoring those from non-automated systems), confirmation bias (i.e., favoring information similar to one's own beliefs or assumptions), ordering effects (i.e., favoring AI recommendations based on the order or timing), and overestimating explanations (i.e., favoring explanations that are considered as high-fidelity) [48]. This is when a thorough, deliberate decision-making process, or *thinking slow* [42], is necessary to minimize potential biases and prevent negative consequences.

We argue that developing and implementing AI-enabled systems requires an understanding and integration of human characteristics, the good and the bad, to foster more favorable use of AI-enabled systems and HAII.

### C. WHY FOCUS ON HAII? Gaps and value of research in HAII

The expected benefits of employing AI-enabled systems, such as improving performance, are well documented in the literature [49]. However, there are too many changing factors playing a role in HAII [19] to be able to realize these benefits unless we study humans and AI-enabled systems together using a multidisciplinary approach [50], [51]. HAII is a complex topic that goes well beyond just the user interface (UI) design [18], [52]. For example, HAII depends on context [39], users [39], and the AI-enabled system itself [2], [5]. This implies that nature and quality of HAII can vary from user to user and context to context even if the same AI-enabled system is being used. This variability is likely to lead to unpredictable outcomes.

[52] suggested several critical concerns for HAII based on the fundamental differences in how non-AI and AI-enabled systems are built. For example, AI-enabled systems can exhibit biased and unexpected outcomes and evolve as they learn, creating non-deterministic output each time. This may lead users to mistrust, distrust, or be unsure whether or when to trust AI-enabled systems [52]. The ethical concerns related to use of non-AI-enabled systems such as privacy, fairness, and decision-making authority are amplified and become much more significant when using AI-enabled systems. [52] has also suggested using a holistic and multidisciplinary approach to tackling these concerns, putting humans in the center of technology.

Although there is growing research into HAII, there is still great variety in how the literature interprets, uses, and describes this interaction. For example, it is unclear whether HAII and similar terms (e.g., Human-System Interaction (HIS), Human Computer Interaction (HCI), Human-AI teaming, Human-AI collaboration, mixed-initiative method, and human-centered AI) refer to the same thing. We chose HAII as the umbrella term to ensure we cover all relevant elements within the topic because it seems to be the most commonly used and general term. This allowed us to capture a greater number of articles, but we believe that a more specific term (or terms) would be better for the field.

In addition, research on HAII is still very immature and inconsistent and does not necessarily focus on the safety-critical industries. This review aims to cover the literature on HAII in safety-critical industries by providing an overview of what has been done and the existing evidence in order to set a more purposeful research agenda.

## III. SURVEY METHODOLOGY

We chose a systematic literature review (SLR) as our survey method, specifically a database search strategy, drawing from our experience reviewing relevant literature rigorously and systematically to minimize bias and produce more reliable results [53]. The SLR focuses on delivering a comprehensive overview of available state of the art evidence within existing literature [54]. It is internationally adopted and accepted as an established and proven method to gather and synthesize evidence in research in areas outside healthcare [54], [55]. In this section, we describe the process used to select, analyze, and interpret a set of relevant articles based on the SLR.

We registered this SLR in Open Science Framework (OSF) [56]. We initially tried to register it in PROSPERO [57], but it was not eligible because our SLR does not include a health outcome with direct relevance to human health and it was not of methodological studies that had a clear link to human health.

Three of the authors (TAB, AB, JKK) have backgrounds in behavioral and social sciences, innovation studies, psychology, biomedical engineering, computer science, and electrical engineering. Two reviewers were female and one male. The lead author and reviewer (TAB) has directed and published previous studies using a SLR with this method [39], [58]. After identifying articles through a targeted search, they were each reviewed independently and separately by each reviewer without revealing conclusions from the other reviewers to ensure consistency and reliability, and to avoid bias. The fourth author (AJ) frequently met with the reviewers to discuss why certain articles were included or excluded and adjudicate disagreements over whether to include or exclude certain articles. We refer to our registered SLR at OSF for more detail on the process of screening and analyzing the articles, as well as reliability and bias mitigation efforts, in greater detail [56].

### A. LITERATURE SEARCH AND STRATEGY

To answer research questions RQ1-4, we established the following search parameters (Table II) based on the inclusion and exclusion criteria in Table I. Only articles that met all criteria were included to highlight evidence available today as well as gaps in the literature.

----INSERT TABLE I: INCLUSION AND EXCLUSION CRITERA HERE----

---------INSERT TABLE II: SEARCH TERM HERE--------

We selected articles to be included in the review following the Preferred Reporting Item for Systematic Reviews and Meta-Analyses (PRISMA) standards [59] (Fig. 1). We searched four large databases to find relevant articles: IEEE Xplorer [60], Scopus [61], ACM Digital Library [62], and Web of Science [63]. We did try to use a fifth database, Semantic Scholar [64], but discovered it had too much overlap with Scopus and/or Web of Science [65]. We chose to use the database search as a strategy because it is efficient, precise, reproducible, and avoids the risk of capturing articles from the same authors multiple times ("identical authors risk" [66]). This was especially important considering that we searched four large databases. This database search strategy was also endorsed by three of our researcher colleagues who have published using a SLR [67]. Based on these considerations, we deemed using the database search strategy as adequate thus not using other methods such as snowballing [68] or hand searching [69].

The search strings and keywords were selected based on the inclusion and exclusion criteria, the research questions, and what we learned from several pilot searches (Table II and Appendix 1). We structured the search strings to cast a wide net while also ensuring that we captured the relevant articles. For example, searching specifically for "human interaction" produced too few articles, but a combination of search strings #3 ("human" OR "humans" OR "user" OR "operator" OR "human factors") and #6 ("interaction" OR "interactions" OR "collaboration" OR "cooperation" OR "teaming" OR "teamwork" OR "integration") captured a wide range of articles focused on HAII. Similarly, we did not search for the full term "safety critical," "industry" or "safety critical industry" because the pilot searches excluded too many relevant articles. We believed this was because the majority of articles relevant to our review did not describe themselves as "safety-critical industries" but rather as their specific industry (e.g., healthcare) or did not mention any specific industry. However, search string #7 (specifically "safety" OR "safe") sufficiently captured articles focused on safety-critical industries or with a safety focus so that we could screen them manually through our inclusion criteria. Additional search combinations like #3 and #2 ("artificial intelligence" OR "AI") were able to capture a wider range of articles than the more limited term such as "Human centered AI" while still including that term. We verified that this method appropriately captured some fundamental examples of the use of neural networks in human interaction, such as chatbots.

### B. THE SELECTION PROCESS OF THE INCLUDED ARTICLES

The literature search resulted in an initial set of 640 articles (Fig 1). After removing duplicates, three authors (TAB, AB, JKK) independently manually screened each article for eligibility by title and then by abstract, leaving 144 articles for full-text screening. After the full-text screening of these 144 articles, the same authors excluded 120 articles because they did not meet the inclusion criteria. For example, articles that were not focused on HAII, used no concrete AI-enabled systems or proof of concept, were not applied in a safety-critical industry, and/or, most importantly, articles that did not involve users (Fig. 1). Group consensus, adjudicated by the fourth author (AJ), was used to resolve any disagreements regarding eligibility. All authors agreed to include 24 remaining articles (3.75% of the initial 640 articles) that fulfilled all inclusion criteria.

--------INSERT FIGURE 1 HERE-------------------------------

We consider the low percentage of articles that met our inclusion criteria to be indicative of the research gap in HAII in safety-critical industries, validating our research objective for this review. However, we have marked the low number of articles as a possible limitation of this review in the Conclusion. Nonetheless, we consider the 24 included articles to be high-quality studies that provide us with an accurate overview of AI-enabled systems currently applied in safety-critical industries with a focus on HAII.

### C. ANALYSIS AND SYNTHESIS

We independently extracted the following information from the 24 included articles: title and citation, author(s), publication year, geographical location of data collection or study, study goal, industry, type of the AI-enabled system, capabilities of the AI-enabled system [70], Technology Readiness Level (TRL) of the AI-enabled system [71], demographic information of users involved, primary role of the AI-enabled system, factors influencing HAII, study methodology, and measurement tool(s). Content analysis was used for data extraction. Common themes were clustered together for each category. Group consensus was used to resolve any disagreements.

*In summary*, we used a combination of digital databases and manual filtering to arrive at the final selection of 24 included articles and annotated them for various variables which we will use to investigate RQ1-4 in the next section.

## IV. OVERVIEW OF THE 24 INCLUDED ARTICLES

This section discusses the 24 included articles (Fig. 2), in which the main findings are summarized in Table III. All articles were published from 2019 through to 2023, with 50% published in 2022 and 2023. Half of the articles conducted studies in the USA and China. Most of the articles conducted studies that aimed to evaluate an AI-enabled system by testing it on their users in a specific use case (75%). Healthcare was the most common safety-critical industry (70.83%).

The dominance of healthcare-related articles might indicate that stakeholders in the healthcare field have a more urgent need to harvest the benefits from AI-enabled systems than in other safety-critical industries. This is likely because healthcare globally has been under great pressure to reduce healthcare workload, improve quality of care and patient safety, increase care delivery efficiency, and reduce medical costs [23], [72], [73] due to healthcare personnel shortages, a growing ageing population, increased healthcare costs, the rise of non-communicable diseases, and emerging disease patterns [74]. In addition, the COVID-19 pandemic amplified those burdens and accelerated the urgent need for healthcare stakeholders to use technology like AI-enabled systems [75].

---INSERT TABLE III: SUMMARY OF MAIN FINDINGS HERE-----------

----------INSERT FIGURE 2: OVERVIEW OF THE 24 ARTICLES HERE----------------------------------------

## V. CAPABILITIES AND MATURITY LEVELS OF AI-ENABLED SYSTEMS USED IN SAFETY-CRITICAL INDUSTRIES

This section discusses the 24 articles on AI-enabled systems used in safety-critical industries based on their AI capabilities and the maturity levels.

### *A. AI capabilities of the 24 AI-enabled systems in safety-critical industries*

Table IV provides an overview of the AI-enabled systems presented in the 24 included articles. A thorough comprehension of the distinctive AI capabilities that enabled the systems to perform their intended tasks was essential for gaining valuable insight into HAII. For this review, we used a comprehensive taxonomy for the European AI ecosystem [70]. The taxonomy clusters AI capabilities into a structured approach to describe the types of basic problems that AI can address and the corresponding use cases in which it can be applied.

-----------INSERT TABLE IV: OVERVIEW OF THE 24 INCLUDED AI-ENABLED SYSTEMS HERE-----------------------------

We sorted these AI capabilities into eight categories with the number of AI-enabled systems under each associated category in parentheses followed by examples from the included articles (Table IV):

1. *Computer vision* (N=9): Making sense of visual data through object recognition and understanding the semantics of visual data. For example, in [76], computer vision assisted radiologists with radiographic interpretations by classifying lesions in mammograms. In [77], computer vision provided real-time vehicle classification and localization to alert highway maintenance personnel of potential risks ahead of time, as well as enhancing robotic interactions and navigation. Further extending the utility of computer vision, [78] and [79] improved diagnostic practices in neurology and pathology, [80] and [81] refined image analysis and annotation, and [77] improved safety measures in dynamic environments. These contributions underscore computer vision's pivotal role across healthcare, safety, and AI interaction domains.
2. *Computer audition* (N=2): Making sense of audio signals (e.g., speech to and from text). For example, in [82], computer audition enabled interactive dialogue between healthcare robots and older patients during a physical exercise session. Similarly, [83] demonstrated how computer audition could enhance interactions in socially assistive scenarios for individuals with Parkinson's Disease, showcasing the technology's adaptability and impact in healthcare environments.
3. *Computer linguistics* (N=7): Also known as Natural Language Processing (NLP); making sense of human languages (e.g., translation, sentiment analysis, text classification, and conversational systems). This domain stands out for its versatility and impact across sectors. From [84], where AI enhances safety in construction through effective communication, to [82], facilitating meaningful interactions between healthcare robots and the elderly, and [85], improving medical diagnostics with descriptive annotations. Additionally, an AI-enabled system's role in creating transparent medical advice in [86], understanding patient needs for self-diagnosis in [87], supporting mental well-being in [88], and clarifying automated driving actions in [89], are a few examples NLP's broad utility in practical, everyday contexts.
4. *Advanced robotics and control* (N=6): Making sense of data representing physical systems to change behaviors (e.g., machine control, autonomous robots). For example, [90] showcased an innovative approach to amplify autonomy within a robotic fleet, specifically using a swarm of self-organizing drones for rescue missions, significantly alleviating the cognitive burden on operators during critical missions. Likewise, an autonomous software system depicted in [91] orchestrated a seamless transfer of control authority among the AI-enabled system, robotic entities, and human users, enhancing task performance and overall HAII. Furthermore, in the realms of healthcare and social assistance, we observed the practical application of advanced robotics and control, notably in [82] where the social robot Pepper engaged older individuals utilizing a blend of robotics, computer vision, and linguistics. Additionally, [92] demonstrated the integration of deep learning techniques in cooperative driving scenarios, where neural networks acted in concert within a dual iterative Approximate Dynamic Programming framework to facilitate synergistic human-automated driving interactions. Lastly, [89] explored an AI system equipped with advanced robotics and control that employs description logic, automated planning, and natural language generation to ensure safe transitions of control in highly automated driving settings. Each article underscores the pivotal role advanced robotics and control play across diverse safety-critical industries, from enhancing operational performance to enriching HAII.

5. *Forecasting* (N=6): Making predictions based on complex patterns (e.g., illness and re-admission predictions). By enhancing predictive capabilities, AI-enabled systems can aid in various domains, including healthcare, automated systems and beyond. For example, [93] focused on surgical risks, and [94] optimizes transfusion planning, and an AI-enabled system's role in mental health and clinical decision support is evident in [95] and [96]. Furthermore, [97] predicted trust based on physiological responses, and [89] forecasted the necessity for manual intervention in automated driving, illustrating the AI-enabled system's broad predictive applications.
6. *Discovery* (N=13): Making sense of massive amounts of data to find patterns such as relationships, similarities, and dependencies (e.g., anomaly detection for cancerous cells). For example, [98] utilized discovery capability to enable Explainable AI (XAI)-driven visualizations of predictions for the presence of breast cancer in patients. This capability is exemplified further in multiple studies across safety-critical industries, for example, enhancing radiologists' performance in lesion classification through algorithmic transparency and interpretability measures [76], [86], improving pathologists' decision-making in prostate cancer pathology [99], and in the development of clinically relevant applications for online Cognitive Behavioral Therapy (CBT) treatments [95]. These articles showcase a broad application of Discovery, from enhancing diagnostic accuracy and treatment efficacy to understanding user trust and system integration [78], [79], [80], [81], [96], [81].
7. *Planning* (N=5): Making sense of long sequences of actions in complex environments for optimal solutions. For example, planning capability was utilized in [94] to provide clinical recommendations to help caregivers determine the number of packed red blood cells required to achieve the target hemoglobin or hematocrit value for adult patients, thereby minimizing the overutilization of blood product transfusions. Planning's role spans further, including [91] for robotic control authority, [92] for collaborative human-AI driving decisions, [90] for swarm robotics strategy, and [89] for safe AI-human handovers in driving, demonstrating planning's broad utility across fields.
8. *Creation* (N=7): Involving new data generation, such as text, speech, audio, and images. For example, [84] used text generation to enhance hazard awareness in the construction industry. Following a similar vein, [85] applied text generation to enable automatic captioning of medical images. Taking it a step further, [82] employed speech generation to foster a more natural interaction between healthcare robots and the elderly. The system described in [86] utilizes creation capabilities to generate understandable explanations for AI-driven clinical decisions, bridging the gap between complex data analysis and human interpretability to aid caregivers and patients. In [87], Creation also aids in developing dynamic, interactive platforms for self-diagnosis, demonstrating AI-enabled systems' role in personal health management. Additionally, [88] explored the therapeutic potential of AI through personalized conversation generation, offering support to individuals facing burnout. Furthermore, [89] illustrates the use of generative AI in enhancing safety and trust in autonomous vehicles by creating clear, timely communications for drivers. These applications all fall under the broad umbrella of what is commonly referred to in the field of AI as Generative AI.

Discovery and Computer Vision capabilities were used most commonly (Fig. 3), highlighting the importance of these capabilities for HAII in safety-critical industries. Additionally, we observed a notable correlation between the use of *computer linguistics* and *creation* capabilities, following the comprehensive taxonomy for the European AI ecosystem [70]. In this taxonomy, "Conversational Systems" are listed under *computer linguistics*, whereas "Text Generation" is listed under both *creation* capability and *generative AI*. It is unclear from the EU taxonomy why these capabilities are separated because, from today's perspective, powerful conversational systems (e.g., chatbots) typically rely both on *computer linguistics* and text generation (*creation*) capabilities. As a result, we have listed the same AI-enabled systems that use *computer linguistics* capability as using *creation* capability as well (N=7) (Table IV). The taxonomy of AI capabilities needs to be adjusted to reflect similar correlations, if any, among the capabilities.

It is also noteworthy that 20 of the AI-enabled systems evaluated in the papers employed more than one AI capability (Table IV), 13 employed two capabilities, five employed three capabilities, and one each employed four and five capabilities. Employing multiple AI capabilities has several advantages for HAII. For example, an AI-enabled system can process and understand inputs from various modalities simultaneously [100] (e.g., it can interpret a spoken command while analyzing visual cues from the environment [89]). This also allows for a more comprehensive and immersive user experience [101] (e.g., users can interact with the system using voice commands, visual inputs, or even gestures), which can make the interaction with AI-enabled systems feel more natural and intuitive [82]. Moreover, multiple AI capabilities enable the system to, for example, analyze the user's facial expressions, tone of voice, and body language to infer their emotions or intentions to gain a better understanding of the user and their context [82], [102].

Additionally, multiple AI capabilities can complement each other in problem-solving scenarios [85]. For example, computer vision can identify objects in an image, which can then be used as input for natural language processing to generate a description [84], [85], [84]. This synergy between capabilities can lead to more advanced and sophisticated HAII.

However, the integration of multiple AI capabilities must be carefully designed to be compatible with a given use case and user profile [103]. It is also important to consider the potential users' cognitive load associated with multimodal systems and whether the chosen modalities are reducing or increasing the cognitive load [103]. Furthermore, this approach may substantially raise the complexity of the system, which can lead to technical difficulties in creating and verifying solutions that are applicable to a wide range of populations [103], [104]. It is therefore important to carefully consider whether and when to use multiple AI capabilities based on the trade-offs to HAII prior to development.

### *B. The maturity levels of the 24 AI-enabled systems in safety-critical industries*

The 1-9 Technology Readiness Level (TRL) scale helps us to understand the maturity of these AI-enabled systems in their intended environment and population (Table IV). However, none of the authors of the 24 included articles assigned TRL levels to their AI-enabled systems. Therefore we identified TRL according to our best judgement following [71].

Knowing the TRL level is important because the environment where the HAII occurs influences the quality and dynamic of HAII [36]. Assigning TRL helps predict how HAII would unfold in real-world settings so adjustments can be made to ensure that the dynamic of HAII is as expected.

We gathered information to assign each TRL from the included articles as well as from providers of the AI-enabled systems when available online. When an AI-enabled system could be assigned into two TRL levels, we chose the lower value to follow a more conservative measure since these AI-enabled systems are applied in safety-critical industries where the level of impact of an overlooked element or a slight misjudgment could cause serious injury [20].

We believe that TRL is an important factor in HAII research and recommend that researchers studying AI-enabled systems should assign a TRL level to their AI-enabled systems and explore TRL implications to HAII more specifically.

In this review, we found that no AI-enabled systems reached TRL 9 (i.e., actual system proven in operational environment). Of the 24 AI-enabled systems presented in the included articles, only four were considered to reach TRL 8 (i.e., system complete and qualified): Telegram (a chatbot) [84], Blood Utilization Calculator (BUC) [94], Pepper (a social robot) [82], and DoctorBot (an intelligent self-diagnosis system) [87]. These TRL 8 AI-enabled systems were the most mature in this review. Of the remaining AI-enabled systems, three were TRL 6 (i.e., technology demonstrated in relevant environment), and one was TRL 5 (i.e., technology validated in relevant environment), 11 were evaluated to be TRL 4 (i.e., technology validated in lab), five were TRL 3 (i.e., experimental proof of concept). None of the articles in our review evaluated AI-enabled systems with TRL 7 (i.e., technology prototype demonstrated in operational environment). Similarly, none evaluated in TRL 2 (i.e., technology concept formulated) or 1 (i.e., basic principles observed), but this was expected following our inclusion criteria to only include articles that presented a concrete AI-enabled system or, at a minimum, a proof of concept.

That only four of the AI-enabled systems are close to the highest TRL level shows that research focused on HAII is still in its infancy in safety-critical industries. This finding may also reflect the fact that all 24 included articles are research papers that focus on testing a certain AI-enabled system to a certain user profile for a specific use case. The nature of research and the specificity it takes are likely the reason most of the included AI-enabled systems are still in the proof of concept and prototype phases.

Importantly though, this finding also shows that users are involved in the proof-of-concept phases. Involving users, ideally as early as possible (i.e., from TRL 1), is an integral part of creating quality HAII by ensuring that AI-enabled systems meet specific user needs and the intended environment's unique characteristics [105], [106]. Users do not need to understand the intricacies of the AI-enabled system's implementation and technical details. However, it is essential to ensure that users are informed about the potential consequences of using it. This is especially important for safety-critical industries where an uncertain situation can escalate very quickly into deadly consequences. Applications of AI-enabled systems in safety-critical industries thus are likely to require a more conservative assessment in assigning the TRL levels, especially for TRL 9 [20].

For future work, researchers may extend the TRL method to consider the quality of HAII during testing. For example, studies focused on users evaluating an AI-enabled system may try to assign a TRL level and explore TRL implications to HAII more specifically. In addition, future work can focus on understanding whether TRL 9 is sufficient to be used in real-world settings in safety-critical industries [20], and whether being TRL 9 guarantees the AI-enabled system will be adopted by users [23].

## VI. RESEARCH QUESTIONS FOR HAII IN SAFETY-CRITICAL INDUSTRIES

This section discusses research questions RQ1-4 for HAII in safety-critical industries.

### *A. RQ1: What terms are used to describe the interaction between humans and an AI-enabled system?*

Table V provides an overview of HAII in the 24 included articles including the terms used, the primary role of the AI-enabled system in the interaction, and factors influencing HAII. Thirteen articles used the terms "interaction" and/or "collaboration" to describe contact between users and the AI-enabled system. One article used "assisting," one used "hand-offs," one used "handover," and one used "partnership." In contrast, seven articles did not use any term to describe the interaction between users and AI-enabled systems. These findings highlight the lack of consensus on terminologies and definitions in the literature.

------INSERT TABLE V: OVERVIEW OF THE FACTORS INFLUENCING HAII-------------

The fact that seven of 24 included articles do not use any term to describe the interaction shows that HAII is likely to cover a broad range of meanings. It can be difficult to use only one term for the interaction between humans and AI-enabled systems due to its complexity and various purposes of AI-enabled systems. Also, choosing a single term potentially restricts the interaction, or it could be distracting or irrelevant to the study focus. This may explain why these articles do not explicitly use a term to describe the interaction. In contrast, seven of 24 articles used both "interaction" and "collaboration," showing that similar terms are likely being used interchangeably (Table V).

The findings also show that there is not one agreed term related to HAII. In addition to the terms "interaction," "collaboration," "handovers," and "hand-offs" that we found in this review, the broader literature contains other similar terms to describe HAII such as "human-AI teaming [51]," "human-AI cooperation" [107], "human-AI symbiosis" [108], "human-AI coordination" [109], and "human-AI complementarity" [110]. All these similar terms may or may not refer to the same thing or have overlapping meanings. Alternatively, the same term may refer to completely different topics. This divergence of terms makes it difficult to look to the research to improve HAII.

The term HAII is often associated exclusively with user interface (UI) design [111], [112]. Whereas UI design is a significant field that enables users and AI-enabled systems to communicate and interact, with the rapid development and increasingly wider application of AI-enabled systems across industries, HAII goes beyond UI.

We use HAII in this review because it is effective and appears to be already the most used term among a wide range of terms throughout the research. Nevertheless, we suggest further research on why these terminologies matter, which terminologies are best in each context, and accurate definitions of the terminology. The goal is for AI communities to be able to come together, discover each other's work, and create a coherent forum for sharing findings that will help others in studying and implementing AI generally and specifically in safety-critical industries. The wide range of terms we found across these articles supports our choice to use broader search words instead of more specific, and limiting, ones.

### *B. RQ2: What is the primary role of the AI-enabled system in the HAII?*

There is not one agreed upon method to categorize the primary roles of AI-enabled systems, so we chose to categorize them into four primary roles based on the degree of their decision-making authority (Table V). This section discusses each role. See Table VI for the descriptions of each of the four types.

---------INSERT TABLE VI: DESCRIPTIONS OF AI PRIMARY ROLES OF HAII--------------------------------------------------------

The most dominant primary role of the AI-enabled systems included in this review was *AI assisted decision-making* systems (N=16). Users still hold the final decision-making authority, but AI-enabled systems can influence that decision-making process. We found that healthcare personnel (e.g., clinicians, radiologists, pathologists) were the users in 15 of the AI-assisted decision-making systems, one targeted patients, and one included users with and without experience working in healthcare. In contrast, only one article targeted construction workers.

In the 15 healthcare cases, it is worth noting that in [76], [94], [113], [93], [85], the impact of the output of the AI-assisted decision-making systems influences not necessarily the users (i.e., healthcare personnel), but the patients who do not interact with the AI-enabled systems. This indirect impact of AI output on people who do not interact with AI-enabled systems may be more prominent in healthcare where the direct users are usually clinicians, rather than in other safety-critical industries where the direct users are also the ones impacted by the AI output such as in [77]. This indirect but significant impact of AI output in healthcare thus likely requires a different lens than the type of HAII we described above. We should consider extending the concept of users from exclusively those who directly use an AI-enabled system to help make decisions and include those who are impacted by those decisions.

In ethical HAII, users are expected to hold the ultimate decision-making agency in the AI assisted decision-making role. However, the process of making a decision with any degree of help of AI-enabled systems [114], can be influenced by cognitive biases and under- or over-reliance and/or trust in AI output [115], [116]. Especially in the context of AI-assisted clinical decision-making, there is still debate over who is liable should a negative consequence happen to a patient: the clinicians, the AI-enabled systems, their developers, or others.

It may be best to focus research on calibrating HAII to foster the appropriate level of trust and/or reliance (i.e., "warranted trust" [117]) in the output of AI-enabled systems during a decision-making process [39], [116]. When designing AI and planning HAII it is important for developers to remember that even though the user has the freedom to ignore the AI's recommendations, it is very difficult to ignore the influence of AI output in clinical decision making [118], [119] once a user is exposed to the AI output [42].

Four articles in our review included AI-enabled systems categorized as *collaborative AI* that require users to work together with AI-enabled systems to achieve a goal [89], [91]. Their users were robot operators (energy) and drivers (automotive). In these articles, AI-enabled systems were treated as equal to the users performing the same task (i.e., navigating or driving), with the goal of getting from A to B safely in different experimental scenarios. In general, AI-enabled systems and users had equal decision-making authority to decide when and whether to hand over control [89], [91], or how to drive the car [92], [97].

Compared to other primary roles (Table VI), *collaborative AI* shares the most equal partnership between users and AI-enabled systems, similar to a description of "human-AI teaming" [51] suggest that teams can be heterogenous (i.e., a mix of humans and AI), as long as the way the team functions is based on coordination and interdependence of all its members in a dynamic setting [51]. This implies that team members need not be equivalent in their "agency, functionality, capabilities, responsibilities, or authority" [51, p.15]. This primary role of AI-enabled systems as an equal team member can become increasingly important in certain situations in safety-critical industries (e.g., military and nuclear energy) in which the threat level to human life exceeds an acceptable level [91]. Nevertheless, for an AI-enabled system to earn such a role requires all stakeholders – regulators, users, and technical experts – to agree that the AI-enabled system in question is indeed ready and safe to take up such role [51].

In contrast to *collaborative AI*, *human-controlled AI* views AI-enabled systems merely as a tool to help users achieve a goal. Two articles in our review described their AI-enabled systems in this way [84], [90]. One system provided users with hazard awareness training [84], whereas the other described AI-enabled swarm systems (indoor micro UAVs) being used to help users on a rescue mission scenario [90]. The AI-enabled systems in this category have very little decision-making authority. This primary role is thus likely to be best suited for applications in very low risk [84] or very high-risk [90] situations where users wish to retain the ultimate control and oversight.

Two articles focused on using an AI-enabled system as *emotional AI* to respond to users' emotions appropriately. Both of these AI-enabled systems were in the form of a social robot. One article targeted mainly patients as users and the healthcare personnel who supervised the AI-enabled system while it interacted with the patients [82]. The second article targeted assisting patients with Parkinson's disease in sorting medications and providing cognitive and social support [83]. *Emotional AI* has very little decision-making authority. Applications of *emotional AI* in safety-critical industries can be even more important because a user's ability to regulate their emotions can be crucial in a safety-critical situation. For example, one article that we excluded in this review described an application of *emotional AI*-enabled systems in the military to identify and respond to, for example, user stress and fatigue prior to entering a critical situation [120]. *Emotional AI* can therefore be used to help manage such situations by regulating and deescalating users' emotions [120].

Future work in AI research needs to explore more of the taxonomy of primary roles in HAII to complete the list and improve understanding. A single AI-enabled system may have several roles in different applications since these roles are context, situation, and application dependent. Additionally, because that role can change the dynamic of HAII, it is likely that in practice the roles are not categorical, but on a scale from 100% human agency (i.e., humans complete a task or achieve a goal on their own) to 100% AI-enabled system agency (i.e., AI-enabled systems complete a task or achieve a goal on their own). When achieving a goal that requires completion of several tasks, it could be that certain tasks are 100% AI-enabled system agency and other tasks are a combination of human agency and AI-enabled system agency.

### C. RQ3: What factors influence HAII?

We categorized factors found to influence the HAII in the included articles into seven categories (Table V).

#### 1) USER CHARACTERISTICS

We found that HAII was influenced by user characteristics including user work position and experience [87], [84], expertise levels [94], [79], personality traits [76], work ethics [81], and user competence and familiarity levels with AI-enabled systems [93], [82], [87], [90], [77]. For example, we found that user's unfamiliarity with AI-enabled systems negatively affected HAII [77], [94]. Understanding the target user's characteristics for specific AI-enabled systems was key to ensure quality HAII. Accordingly, designing AI-enabled systems to fit certain user characteristics that were less likely to change (such as personality traits, and expertise levels) might be key for user acceptance. On the other hand, targeted training and education might be more suitable to improve user competence and familiarity levels with AI-enabled systems, characteristics that were more likely to be changed [93], [77].

Notably, the users involved in the 24 included articles were all end-users following our inclusion criteria, meaning users who directly interact with the AI-enabled systems. In general, the users were not experts in AI-enabled systems, but most were domain experts in a safety-critical industry (Table VII). The number of users involved in the included articles ranged from four to 129 users (Mean=26.67). Not all articles showed the age of users and percentage of female users. For example, only 10 articles reported percentages of female participants, ranging from 10% to 84 %, and one article reported only partially [87]. Six articles reported the age of their users, ranging from 25.13 to 39 years old, and one article reported only partially [87].

#### 2) USER PERCEPTIONS AND ATTITUDES

We found that the following user perceptions and attitudes influence HAII: 1) user considerations to adopt an AI-enabled system; 2) user biases and the fear of being replaced by AI-enabled systems; and 3) user attitudes towards AI-enabled systems. This section will discuss the findings in this order.

a) **User considerations to adopt an AI-enabled system**. The included articles were all research-based and focused on user evaluation of specific AI-enabled systems. They found that even if users interacted well with the AI-enabled system being studied, the users might have other considerations about whether to use it in real-world safety-critical industries [81], [76]. For example, medical practitioner users reported that evidence of an authority approval, such as the FDA, published validation in peer-reviewed journals, social endorsement by perceived leaders in the field, perceived impact of existing workflows and legal liability, and cost of purchasing the AI-enabled system, all influenced their decision whether to purchase and use the AI-enabled system in real-world settings [99], [81]. Integrating AI-enabled systems being studied into daily routines or workflows was also reported to be integral for real-world adoption [81]. For example, the patient users of the AI-enabled system robot in [83] reported that the procedure for medication sorting used by the robot was different than how patient's usually sorted their medication (which also varied from patient to patient), creating confusion and doubts about using the robot daily. These findings show that there is more to consider in bringing an AI-enabled system from bench to bedside than simply ensuring users to use AI-enabled systems, especially in real-world safety-critical industries where the stakes are high.

b) **User biases and fear of being replaced by AI-enabled systems**. All these adoption considerations might be, to a certain extent, rooted in users' fear of being replaced by AI-enabled systems [86], and possible user biases [95], [96], [79], [81], [98]. For example, in [98], users were found to trust the AI-enabled system more when it provided a diagnosis as malignant rather than benign, showing a bias towards a malignant diagnosis. Another example, user over-reliance to AI-enabled systems [95], [88], might happen in a situation that involves a fast-paced and overwhelming work environment, typical in safety-critical industries, such as a busy hospital where users deal with complicated conditions or cases [95]. In these situations, users may default to accepting the recommendations by AI-enabled systems instead of their own training especially when the AI-enabled system produces a high confidence levels [95], [88].
Fortunately, users generally acknowledged and were mindful that AI-enabled systems could make errors [95], [79] and would thus use AI output as additional information for guidance rather than authoritative [95]. This was reflected in [97] where users were found to rely on the AI-enabled system significantly more during no-risk driving conditions than during very-high-risk and high-risk driving conditions. In addition, users reported that they would immediately become less trusting of the AI-enabled systems when they encountered incorrect outputs [95]. Users were also reported to prefer false positive rather than false negatives because they perceived the consequences for false negatives to be much greater [79]. There is a continuous balancing act between fostering user trust and convincing them to use AI-enabled systems on one side, and ensuring users remain mindful of AI-enabled systems' imperfections on the other. This might be achieved by, for example, deliberately letting users reach their own conclusions prior to exposing them to AI output [96], targeted AI training and supervision [95], and presenting both AI capabilities and limitations (See the Explainability and Interpretability section).

c) **User attitudes towards AI-enabled systems.** We found that user attitude towards any AI-enabled systems generally determined how users would react to a specific AI-enabled system [88], [76], [87]. For example, if a user mistrusted AI-enabled systems in general, the user was likely to view benevolence of an AI-enabled system as manipulative [88]. In contrast, a user who has a very positive opinion of AI might over-rely on an AI-enabled system [88]. Users' positive attitude towards specific AI-enabled systems was reported to be linked to perceived reduced workload when using an AI-enabled system to complete a task [79] as well as perceived high accuracy, reliability, and functionality [77], especially for cases that were difficult for human experts such as diagnosing rare diseases [87]. However, [97] found that attitudes could change and improve and that trust in AI-enabled systems was dynamic. Additionally, user preferences such, as design choices, were found to influence HAII [90]. For example, the fact that certain XAI methods were perceived as more user friendly than others [78].

### 3) USER EXPECTATIONS AND EXPERIENCE

We found that the following user expectations and experience influence HAII: (1) mismatch between user expectations and experience, (2) conflicting AI output and user conclusions, (3) user perceived value of AI-enabled systems.

a) **Mismatch between user expectations and experience.** Individual users were likely to have different expectations prior to interacting with an AI-enabled system that might or might not match their experience. Actual user experience interacting with the AI-enabled system [97], AI output [95], [99], behaviors of AI-enabled systems [83], and AI interface and features [83], were all reported to be factors that users usually compared to their expectations. For example, in [99], some medical practitioner users expected AI output to accurately predict outcomes, whereas others expected to use AI output to merely draw their attention to suspicious regions thus were less bothered by inaccuracy. In another example [83], patient users expressed that the AI-enabled system robot's management of medications did not match their usual management. The patient users were also concerned by the AI-enabled system robot's speech, shape, and size [83]. Such unmet expectations during interaction easily led to user distrust [99], frustration [83], and disuse [83], especially if the user attitude towards AI-enabled systems in general was already negative [96], [81]. Expectations could be unmet because the AI-enabled systems were tested in a different or a more general population than target user groups [83], [86] (such as students instead of patient users [83]) or because users received significantly different information about the AI-enabled system's capabilities and limitations than the reality [97].

b) **Conflicting AI output and user conclusions.** Users often expect AI-enabled systems' output and recommendations to be similar to their own conclusions [94]. They respond positively when their conclusions match AI output and negatively when they do not [76], [80]. Understanding how to verify AI output and AI performance helps resolve this conflict. [99] suggested using AI performance metrics that users are familiar with to avoid confusion, while [78], [99] suggested presenting AI rationale (i.e., XAI) in a way that closely resembles users' own decision making processes. A few studies, [96], [79], [80], suggested that understanding users' thought processes when making decisions is crucial. For example, when healthcare personnel used AI output to see whether a clinical treatment had any effect on patients, they would report a positive result if the output matched their expectations [95]. However, when the output showed the treatment had no positive effect, users' reaction could be either constructive or destructive. Constructive reactors explore ways to understand and find explanations by engaging patients and looking other clinical treatments or motivating themselves to advance their own competence. Destructive reactors attribute the limited effect to external factors or their own incompetence [95]. A calibration phase between AI recommendations and user conclusions against the ground truth (i.e., expert judgement) could settle conflicts between them [99], [96] and show users the value of AI-enabled systems [79]. This was especially true for making a clinical decision [79]. These findings highlight the importance of aligning user expectations to their experience.

c) **User perceived value of AI-enabled systems**. Users expected that using an AI-enabled system would bring value to them. Users perceived value of AI-enabled systems was usually based on their perceived AI accuracy and perceived user friendliness [92], [94], [79], as well as when users felt they were able to perform better (e.g., more efficiently and effectively) with the AI-enabled system's assistance and when they believed that using the AI-enabled system was better than not using it [79], [81], [92], [76]. Improving user perceptions of the value of AI output could be achieved by using user feedback as input data to train AI models on the fly [81]. We found that it is therefore important to motivate users and provide immediate incentives to provide feedback or input data [80], [88].

### 4) AI INTERFACE AND FEATURES

The key finding here is the importance of personalizing the design features and user interface of the AI-enabled system according to the users' needs and preferences. For example, how an AI-enabled system communicated and interacted with users [91], [87]. In [91], users wished to have a more direct and immediate influence on the decision whether to take over control authority from the AI-enabled system in highly automated driving.

Another factor was which design features were prioritized for which users [98]. For example, some users preferred to have more interactive displays or interfaces so that they could make changes, and some other users preferred to have additional or prioritized information readily accessible [113], [76], [95], [79]. One way to personalize designed features to users was a progressive disclosure of information or a hierarchical design [79]. This means that users were not presented with all detailed information immediately, but could request progressively more detailed information from a high-level overview to the most granular information (e.g., such as a complete documented patient history) [81], [79]. Such features might also include a function to hide or keep certain information on display while requesting for more and more detailed information [81], [80], [79]. This kind of personalization design might prevent redundant and unnecessary information from overwhelming users while keeping relevant, desired information on display [80]. Providing users with more control and shortened the time it took to interpret AI output, especially if their task was time-critical, such as in a medical setting [79].

Where feasible, users preferred simpler and clearer visualizations where possible instead of numbers and texts or narrative presentation [95], [86], [78], [93], for example, in the form of an easy to interpret heat map [78], [79], [81], [76] or even in a table and bar charts [93]. The use of gradation or colors to draw users' attention and filter out information can reduce information load for users [80], [81], [93]. In general, users also desired content on the AI interface to be quick and easy to read and see [79], simpler, and contain more precise information [86].

Importantly, AI emotive response capabilities (e.g., empathy) were considered significant for AI-enabled systems whose task was to respond to user emotions, and sensitivity capability was considered significant to understand the user's complex emotions [87], especially in [82]. The lack of emotive response features in the AI-enabled system (i.e., being in the form of a robot) was a significant drawback in this case because the users were elderly patients and preserving their dignity was considered to be significant part of care and a defining factor of whether patients felt comfortable receiving care or interacting with the robot [82].

Finally, incorporating features in the AI interface for users to provide feedback on the AI-enabled systems were seen as important and desirable [88]. For example, a feature for users to click whether they approved, declined, or were uncertain of AI output was suggested as a way to calibrate AI and improve output and user conclusions [79].

#### 5) USAGE OF AI

How an AI-enabled system was used by users determined HAII. For example, the effectiveness of an AI-enabled system increased when used in a less complex scenario, and decreased in a more complex one [84]. Similarly, users relied on AI-enabled systems more frequently in situations when risks were non-existent compared to high-risk and very-high-risk situations [97].

In addition, we found the heterogeneity of AI-enabled system applications, environments, user needs, and how and when users use it to be important [94], [93]. An AI-enabled system might be used in various environments and embedded in various activities that the developers did not consider. This occurred in [94], where the AI-enabled system had the opportunity to be used on patients whose profiles were outside its scope. Tailoring UI to automatically detect the patient population being treated is a feature that would avoid the AI-enabled system from being used outside the target patient population [94]. This also occurred in [80], where pathologist users used a prediction map, originally developed for users to assess AI model's performance, to help them to see whether there was something more worth looking closer or to double check their potential diagnosis. In contrast, AI-enabled systems might become unusable if developed for a very specific or specialized task [81].

#### 6) EXPLAINABILITY AND INTERPRETABILITY

Understanding user needs was important to ensure the information the AI presents to the user is not excessive, unnecessary, or confusing [86], [96]. Specifically, [96] suggested presenting individual aspects of AI rationale as a single cohesive explanation that matched the user's needs is an effective way to reduce information overload and improve HAII [96].

Explainability and interpretability factors might include the level of detail of output, user level of understanding of the algorithms used [98], logical reasoning and reliability of the AI-enabled system, the context of the AI rationale, traceability of evidence leading to AI output [79], and the speed this information could be understood and interpreted by users [93], [86], [95]. For information that might not make sense if taken out of context, [96] suggested presenting users with supplementary information about context so they could verify the AI output and build trust.

[99], [79] found that users preferred when the AI-enabled systems' strengths, pitfalls, and limitations were presented alongside its capabilities because it improved HAII by allowing users to anticipate these characteristics when interpreting AI output. Presenting this information would also help match *user expectations to experience*. Helpful information about limitations could include: whether the AI-enabled system corrected itself prior presenting AI output to users, whether the AI can determine when it is wrong, whether the AI was trained on the same kind of inputs the user was provided with, whether the data used for training algorithms came from respected and credible sources, and whether the AI-enabled system was based on more reliable information than what the users had access to [99]. AI rationale processes might need to be designed to mimic users' processes in specific settings so that users could verify AI output and reduce perceived uncertainty, especially when AI output conflicted with users' own conclusions [99], [79].

[98] found that when users did not receive explanations (and therefore did not understand AI rationale processes) it degraded their trust in the AI-enabled system. However, although explanations increased user understanding of the algorithms and AI rationale, it did not guarantee user trust. Similarly, [95] and [96] found that explanations of AI rationale were critical for user acceptance of AI-enabled systems because it allowed users to experiment with the systems and verify AI output. It is therefore valuable to test different ways of explaining AI rationale with users to find the best fit for each user profile [78], [79].

#### 7) AI OUTPUT

Determining how and which AI output information is presented to users [94], [93], in which format and timepoints [93], [89], in what order [81], [93], and in what manner [93], [89], [79], [81], influences HAII. Therefore, user expectations and needs should be considered.

Based on our findings, AI output that includes the following information could help users to decide whether to trust and use the output information: certainty [76], validity [76], accuracy [87], [98], reliability and confidence levels [80], [93], the AI models' potential failure points and limitations [99], explanations, additional justification for negative output [80], and both the focus and context when presenting AI output with multiple criteria [79]. Understanding AI-enabled system potential weaknesses would help users to compensate for them while interpreting and verifying the output, and could prevent over-reliance [99], [79], [81]. [79] recommended another way of preventing over-reliance is to encourage users to make their own decisions prior to receiving AI output.

Importantly, users valued receiving actionable recommendations and suggestions, as well as possible implications of these recommendations [99] and evidence leading to these recommendations [80], to help them make decisions [113], [93], [80]. Users also preferred an option for more detailed information when they desired it or automated personalized information tailored to individual users [93], [93], [89]. When multiple AI-models needed to be used, AI-enabled systems needed to integrate results from these different AI-models in an interconnected way and present one final, unified AI output rather than presenting each AI model's output separately [79].

We also found conflicting findings between two articles that measured whether the AI-enabled systems made users work more efficiently [79], [81]. In [81], users were 35% more efficient when using their AI-enabled system. In contrast, in [79], users took 1 min 17 seconds longer when using their AI-enabled system due to the extra workload to interpret and verify AI output.

--------INSERT TABLE VII: OVERVIEW OF THE PARTICIPANT DEMOGRAPHIC INFORMATION, DEPENDENT VARIABLES, AND MEASUREMENT TOOLS AND METHODS HERE---------------------------------------------------------------

None of the 24 included articles covered all seven HAII influencing factors (Table V). Most of the included articles listed user perceptions and attitudes (N=16) as a factor influencing HAII, followed by user expectations and experience (N=14), AI interface and features (N=13), and AI output (N=13). Critically, these seven factors are interconnected and need to be improved simultaneously to ensure quality of HAII. Providing users with more detailed information on the rationale of AI output (i.e., improvement in AI output and explainability and interpretability) is likely to improve user perceptions as well, and effort to improve user perceptions is likely to also improve the content and presentation of AI output. Therefore, we propose a sequence for addressing the factors to optimize the quality of HAII and to emphasize that the process of improving HAII is continuous (Fig. 4).

------------------INSERT FIGURE 4 HERE --------------------

*User characteristics* as well as *user perceptions and attitudes* should be explored first to understand the specific needs of users. This is an essential first step of any AI (development) project. Once they are understood, the information can be used to design the AI-enabled system's interface and to prioritize features that are important to the users (i.e., *AI interface and features*). In addition, exploring how the AI-enabled system can be used in ways other than anticipated, and which situations are best and worst, ensures appropriate use (i.e., *usage of AI*). It is worth noting here that understanding how the AI-enabled system will be used should also include the operational environment where it will be implemented and used. The same users may interact with an AI-enabled system differently if put in different environments. A use case is therefore important for anticipating how HAII may happen in various real-world settings.

Once a use case or cases are settled, and possible uses of the AI-enabled system are explored, the next focus is to ensure *explainability and interpretability*. In the literature, these terms have been used both interchangeably [121] and given explicitly different definitions [122]. In this review, explainability and interpretability are both part of ensuring that miscommunication is minimized between the AI-enabled system's explanation of its rationale and what the users understand about its rationale [122]. Explainability here means that the rationale behind an AI-enabled system's decision is being explained to users in the most efficient and effective manner at the right time. Interpretability here means that users can understand what is being explained correctly, accurately, and in a timely manner.

Similarly, presenting *AI output* to users can be done in a specific manner and time point as desired by users. The difference between AI output and explainability and interpretability is that the former focuses on how AI output is delivered and explained to users, whereas the latter focuses on explaining the thinking process (the rationale) prior to producing an output. They are both focused on closing the gap between what is being explained to users and what they understand. In other words, minimizing miscommunication and misinterpretation.

To match *user expectations to user experience*, it is important to test the AI-enabled system with users, ideally starting in the earliest stages of development and throughout the process from ideation through to deployment [83]. Unfortunately, most often user testing only happens once the development of an AI-enabled system is completed [86]. Testing with users can also confirm whether the initial idea of user profiles (the user's characteristics, perceptions, and attitudes) matches the AI-enabled system being tested [86]. Involving users in testing the AI-enabled system is best seen as an iterative process from development through to deployment and monitoring [86], [96].

### *D. RQ4: How is HAII being measured in safety-critical industries?*

17 of the 24 included articles (70.83%) used an experimental study either as a standalone study methodology or in combination with qualitative (i.e., interviews) and/or quantitative (e.g., a questionnaire) methods (Table VII). Specifically, they used users' subjective measures (i.e., data collected from human responses to questionnaires or human observations and interviews [123]) and objective measures (i.e., data collected from physiological and physio-psychological measures [123], as well as outcome measures) as the outcome variables.

16 included articles used only subjective measures as outcome variables (Table VII). Examples of subjective measures are user evaluation or perceptions of the AI-enabled system [76], [94], [87], [77], [98], user trust [94], [85], [87], [89], [94], user preferences in design choices and handover methods [93], [113], [89], user perceived workload [91], [90], , user performance [76], user perceived potential impact of the AI-enabled system on clinical workflow [93], user situational awareness [84], user attitudes [76], [88], user personality [76], and the level of agreement with AI output [93].

These findings show the importance of users' opinions on a range of factors of AI-enabled systems and thus consistently highlights the importance of involving users in any development phase of AI-enabled systems. End-user development (EUD) [105], for example, can be a suitable method for involving users in tailoring an AI-enabled system at any point in the development phase and beyond.

Seven included articles used both subjective and objective measures (Table VII). For example, measured user workload was compared with perceived workload [91], [90]. Another example, measured user task completion time was compared with user perceived task difficulty [78], or measured efficiency (time saved) was measured together with user perceived effort and workload [79]. Examples of objective measures used were: user task completion time [91], [78], eye movement [84], and heart rate variability [90], AI actual accuracy [78], [79], and efficiency [81], [79] (Table VII). There is only one included article that used only an objective measure (i.e., eye movement) [84].

Objective measures could provide valuable information about HAII that users might not necessarily be aware of, even though their results might not always align with results of subjective measures. This was the case in one included article that measured user cognitive load subjectively and objectively [90]. Their finding showed that a tangible control interface (vs. a digital one) was perceived by users as cognitively more demanding (i.e., a subjective measure). However, results from an objective measure (i.e., heart rate variability) showed that users' cognitive load was not increased. We also found opposing results of subjective and objective measures in another study in the literature that investigated whether cognitive forcing functions can reduce user overreliance on AI-assisted decision-making systems [115]. This study found that users' most preferred and trusted AI-assisted decision-making systems were also perceived as less demanding by subjective measures, but those same systems had worse objective user performance. In another included article [91], the results from objective and subjective measures were in agreement that their proposed method for control authority transfer was better than a baseline method. However, in this case the subjective and objective tests measured different variables than in [90]. Similarly, another included article showed that compared with a manual driving, user cognitive workload was reported to be lower, both objectively and subjectively, compared to driving using an AI-enabled system [92].

We observe that 16 out of the 24 included articles (66.67%) use solely subjective measures, potentially because these measures are more valuable in understanding users' buy-in and predicting what factors are important for users when interacting with an AI-enabled system. Subjective measures can also provide information about which influencing factors are more prone to change and can therefore be improved. For example, user acceptance, user perceptions, user trust, and user design preferences are all subjective and more easily changed or improved than, for example, user personality. Understanding users' subjective needs requires subjective measures so that developers can create an environment where users are willing to interact with an AI-enabled system and ensure quality HAII.

This does not mean that objective measures should be abandoned. Rather, our findings show that there is a research gap in direct and objective HAII measurements. Subjective measures depend on the user's knowledge, memory, and self-reflection. Objective measures, which are usually based on physiological, physio-psychological, and outcome measures (e.g., task completion time, number of collisions), can give rich information that cannot be gathered otherwise. The question is not whether subjective or objective is a better measure, but how to benefit from both to paint a more complete picture of HAII. Combining both objective and subjective measures can be beneficial to verify results from each method and explore a phenomenon more thoroughly than that of using one method, showing that both methods are complementary [123].

## VII. CONCLUSIONS

There are many factors that influence HAII. Effective AI must be based on a well-researched use case that understands the users and their environment. One size does not fit all when it comes to ensuring quality HAII in safety-critical industries. Whereas it would be nice to be able to develop a single AI-enabled system that would solve a wide range of problems, our research here shows that AI and HAII are far too diverse for that to be possible. In an ideal world, AI would be able to detect the user's characteristics and tailor itself accordingly. However, we found that the lack of consistency in the research and the wide range of AI applications and users mean that researchers and developers still need to tailor their AI-enabled systems to their users.

Many of the articles in the literature did not meet our inclusion criteria because they did not involve their target end-users in their studies. Based on our findings, this is a critical factor for ensuring quality HAII that may often be overlooked. For example, one included article that aims to develop and evaluate an AI-enabled system in the form of a social robot to assist older patients with Parkinson's disease reports that the results are mainly positive when testing the robot on students [83] but face many unforeseen challenges when the real target users get to try it out. Our findings can serve an explanation on why involving target users is crucial. The fact that users' characteristics, backgrounds, expectations, perceptions, and attitudes, among others, are found to influence HAII, shows that variations are most likely to happen when the same AI-enabled system is being used by different user groups. We thus argue that HAII research and user evaluation of AI-enabled systems should always involve target end-users.

The environment where an AI-enabled system is to be used can have a significant influence on HAII. For example, the four TRL 8 AI-enabled systems included in this review are commercial products that can be applied in other settings than the ones described in their articles. It is likely that a different application in a different setting will create a different dynamic of HAII.

The substantial gap in HAII research is not only because there is still too little research in HAII, especially in safety-critical industries, but also because HAII is a complex and multidimensional research field, which covers different branches of disciplines and demands a multidisciplinary approach. Having said that, the conclusion that AI needs to be tailored to its users is not unexpected considering the level of complexity that AI-enabled systems and humans bring to the table. Specifying a use case is therefore beneficial and may be the only option for creating successful HAII. Accordingly, future work in AI development and research needs to explore how the unique characteristics of each safety-critical industry influence HAII and how cultural aspects may influence HAII in safety-critical settings. For example, one study found that East Asian participants are more likely to project trust towards Emotional AI systems than Western ones [124].

Another limitation we found in the research is the seven factors that influence HAII (Fig. 4) do not include how long or how often the interaction happens. This is likely because typical research on HAII focuses only on a limited time stamp or a one-off interaction. Nevertheless, the frequency and length of HAII should be explored for its possible influence. One study that did not meet our inclusion criteria found that a longer-term relationship with HAII was perceived as a key element of maximizing the benefit of AI-assisted decision making systems [114]. Similarly, findings from a study on user trust in AI-enabled systems suggest that user trust increases over time along with familiarity between users and AI-enabled systems [39]. The hypothesis that longer or more frequent HAII can be beneficial may be related to AI-enabled systems' ability to learn. Thus, more interaction means more familiarity, knowledge, more mature adjustments, and more fine-tuning regarding how the AI and users work together. Future work thus should explore how HAII can be developed into human-AI relationships, extending the focus from a short-term or one-off contact into a longer-term interaction or relationship in safety-critical settings. Specifically, researchers should investigate the benefits and trade-offs for HAII if humans and AI-enabled systems are given a chance to interact more over a longer term. Importantly, more research is needed to discover how the seven influencing factors found in this review can be used to improve HAII.

There are some limitations in this review. First, only 24 articles (3.75% of the initial 640 articles) met our inclusion criteria. This may be seen as relatively limited and our findings should be generalized with caution and should not be taken out of context. However, we feel that these 24 articles faithfully exemplified the current research in the field and the small number should be seen as indicative of the need for further research on this topic. Future studies may be improved by implementing a snowballing [68] or a hand searching methods [69] to explore an additional perspective of the literature. Second, this review does not include grey literature, which may contain other relevant or more recent work on HAII. Similarly, there could be other relevant articles that are not included in the databases used in this review. Third, each safety-critical industry may have unique properties with implications for HAII. We did not explore these possible implications in this review because our goal was to focus on commonalities across industries, rather than differences between them. In addition, there are only one to three articles for each industry included here except for healthcare (Fig. 2), which is too small of a sampling to draw conclusions about differences between industries. Fourth, findings may be skewed because healthcare as an industry, and USA and China as study locations, dominate the findings.

In conclusion, we argue that HAII in safety-critical industries should be the focus of any context and situation involving an AI-enabled system. In fact, we argue that the development of AI-enabled systems must involve users at all stages. No matter how sophisticated and safe an AI-enabled system is, it cannot achieve its potential unless users are able to interact with it as intended in the intended operational environment. A collective, multidisciplinary effort is required to unpack and understand how to unlock effective HAII in safety-critical industries, beginning with a thorough investigation of the terms and definitions.

## APPENDIX

-----INSERT APPENDIX 1 HERE ----------------

## ACKNOWLEDGMENT

We would like to thank Dr. Tobias Gauss, MD, for his feedback on an earlier version of this manuscript.

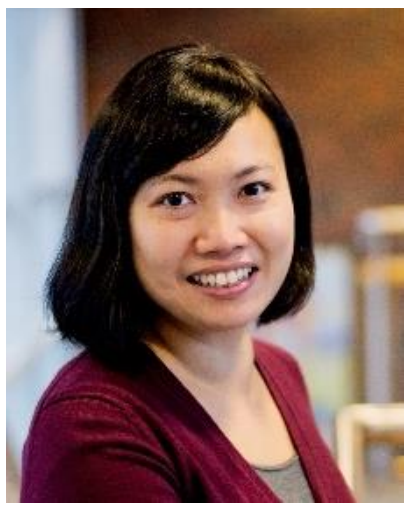

**TITA A. BACH** received a PhD in behavioral and social sciences from the University of Groningen, Groningen, the Netherlands, in 2012, and a master's degree in health psychology from Leiden University, Leiden, the Netherlands, in 2007. She received a diploma in Top Tech Executive Education from Haas School of Business, University of California, Berkeley, CA, USA, in 2017.

In January 2023, she joined Digital Transformation Team, Digital Assurance, Group Research and Development, DNV, Høvik, Norway, as a Principal Researcher. From 2011 to 2022, she was a Principal Researcher in Healthcare programme at the same organization. Her research interests include human behaviors with technology in an environment, organizational psychology, user trust in technology and AI, AI deployment and adoption, safety culture, and cybersecurity culture.

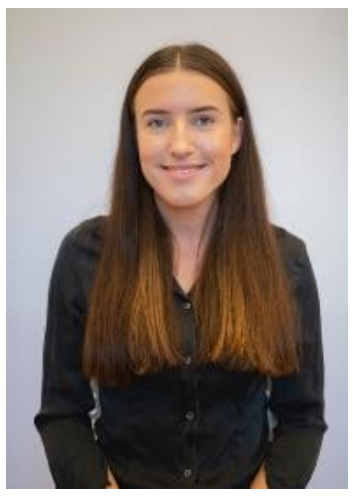

**JENNY K. KRISTIANSEN** received a master's degree in innovation studies in 2022, and a bachelor's degree in psychology in 2019, both from the University of Oslo, Norway.

In October 2022, she joined the Digital Transformation Team, a part of the Digital Assurance research programme in Group Research and Development in DNV, Høvik, as a Researcher. Her research interests include cognitive neuroscience, responsible AI, and human biases in decision-making and its potential influence on the development and deployment of technology and AI.

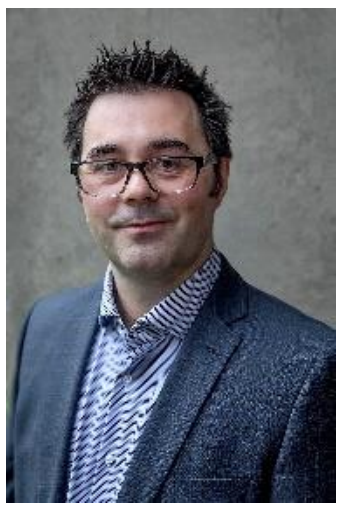

**ALEKSANDAR BABIC** obtained his Ph.D. in biomedical engineering from the University of Oslo, Norway in 2019, and his M.S. degrees in computer science and electrical engineering from the University of Belgrade, Serbia in 2012 and 1999 respectively.

He is a Principal Researcher at the Healthcare Programme, Group Research and Development of DNV in Høvik, Norway. He has over two decades of industrial R&D experience and has worked on various projects such as machine vision for 3D cameras and image fusion and deep learning for the applications of cardiac ultrasound. His research interests focus on the challenges associated with the implementation of AI in healthcare, including developmental and regulatory aspects, trustworthy and explainable AI, data quality and access, privacy, and algorithm development and robustness.

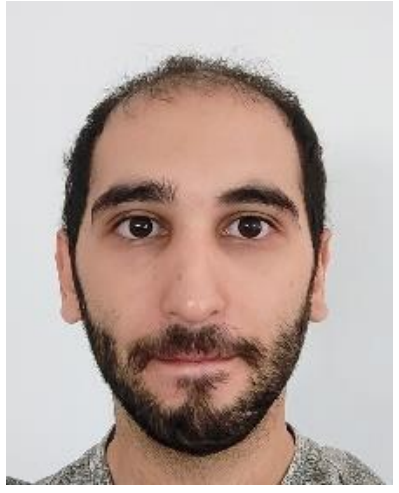

**ALON JACOVI** is a PhD candidate at Bar Ilan University under the supervision of Prof. Yoav Goldberg in Israel. He received his B.S and M.S degrees in computer science at Bar Ilan University as well, in 2017 and 2019 respectively.

His work primarily involves explainability and interpretability in natural language processing, as well as investigating trust in AI.

Mr. Jacovi is a member of the Association of Computational Linguistics community and has completed research internships at IBM (2016-2020), RIKEN (2019), AI2 (2020) and Google (2021-2023).

**List of Figures**

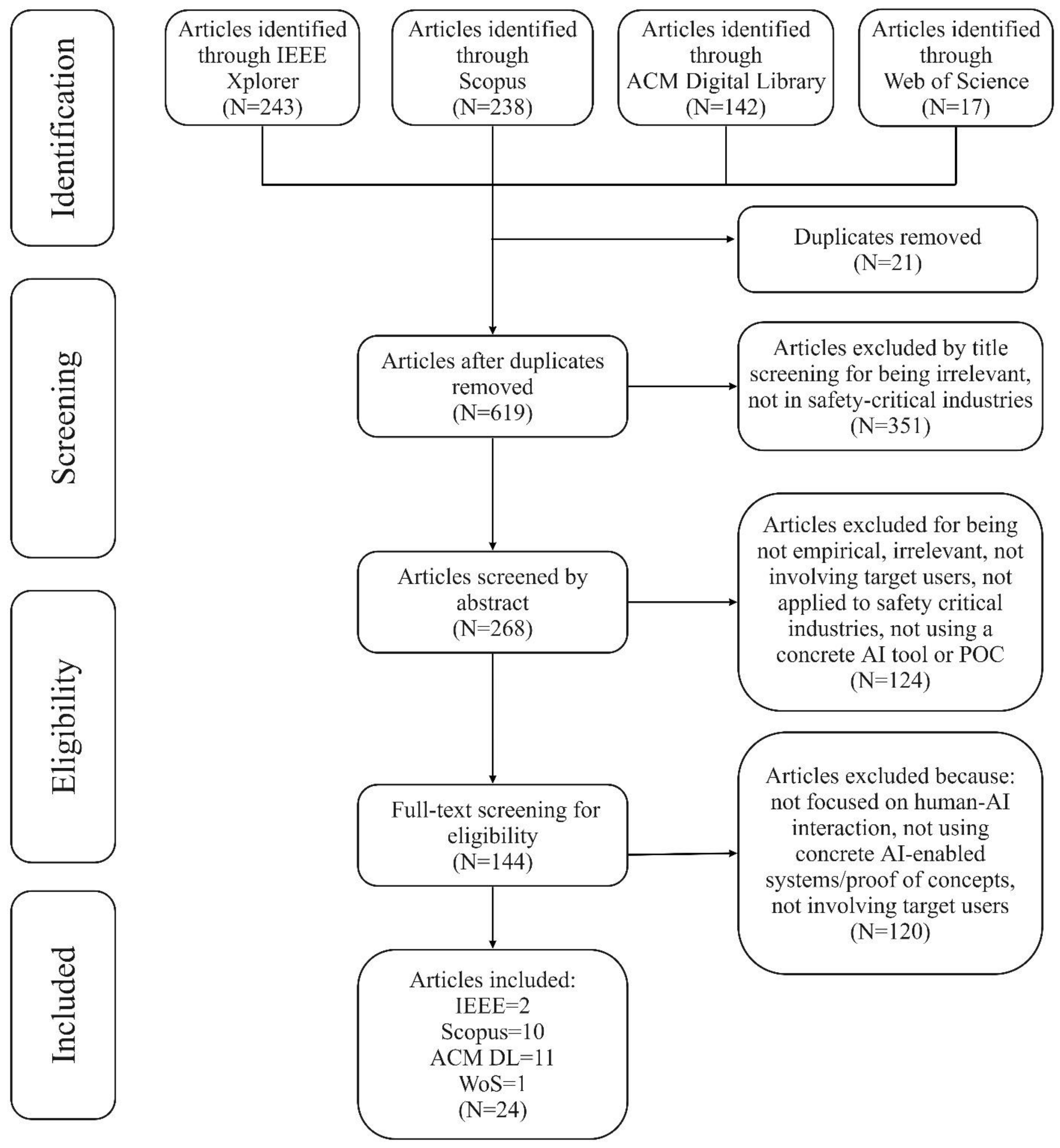

**FIGURE 1.** PRISMA flow chart showing the selection process and the number of included and excluded articles at each stage

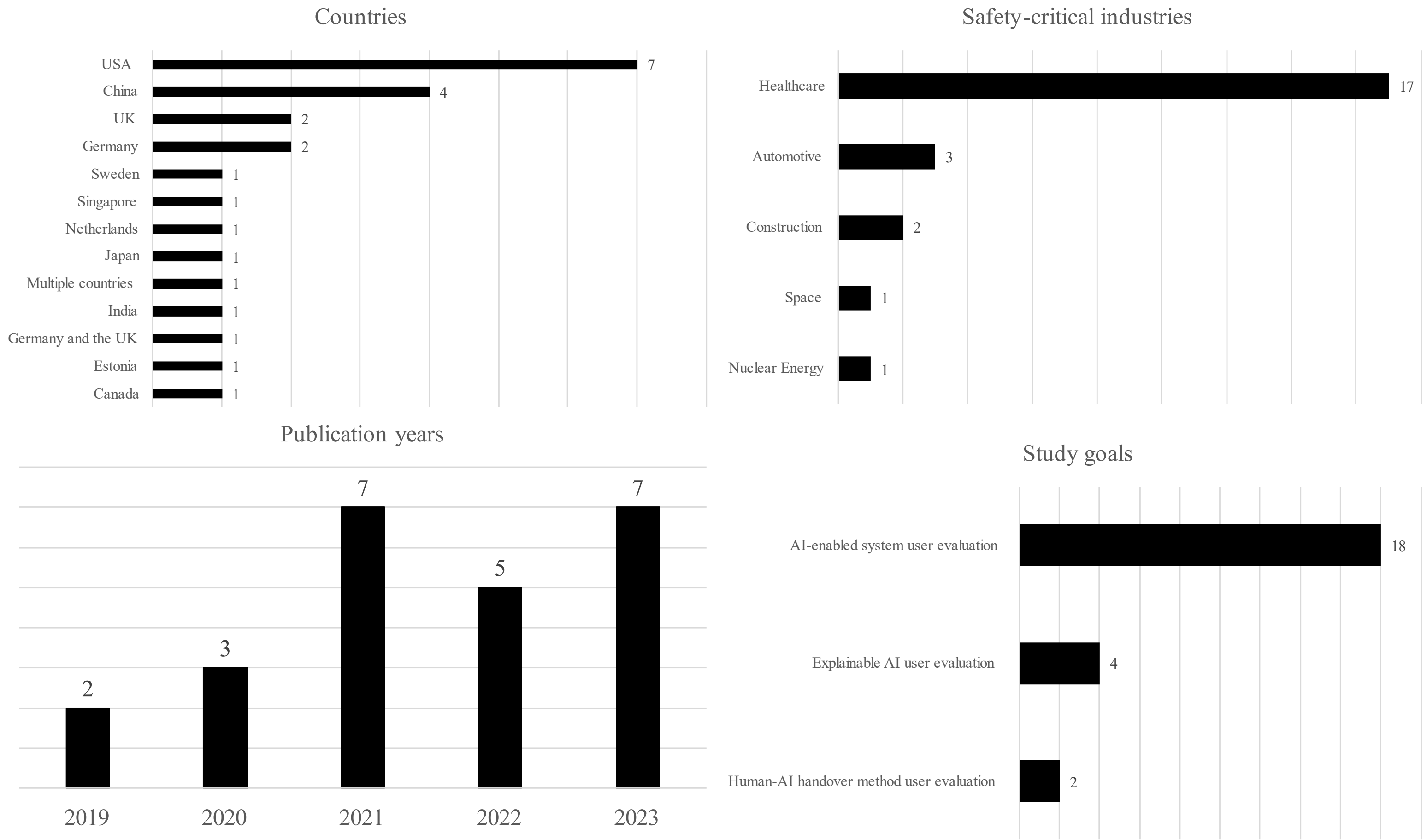

**FIGURE 2. Overview of the 24 Included Articles**

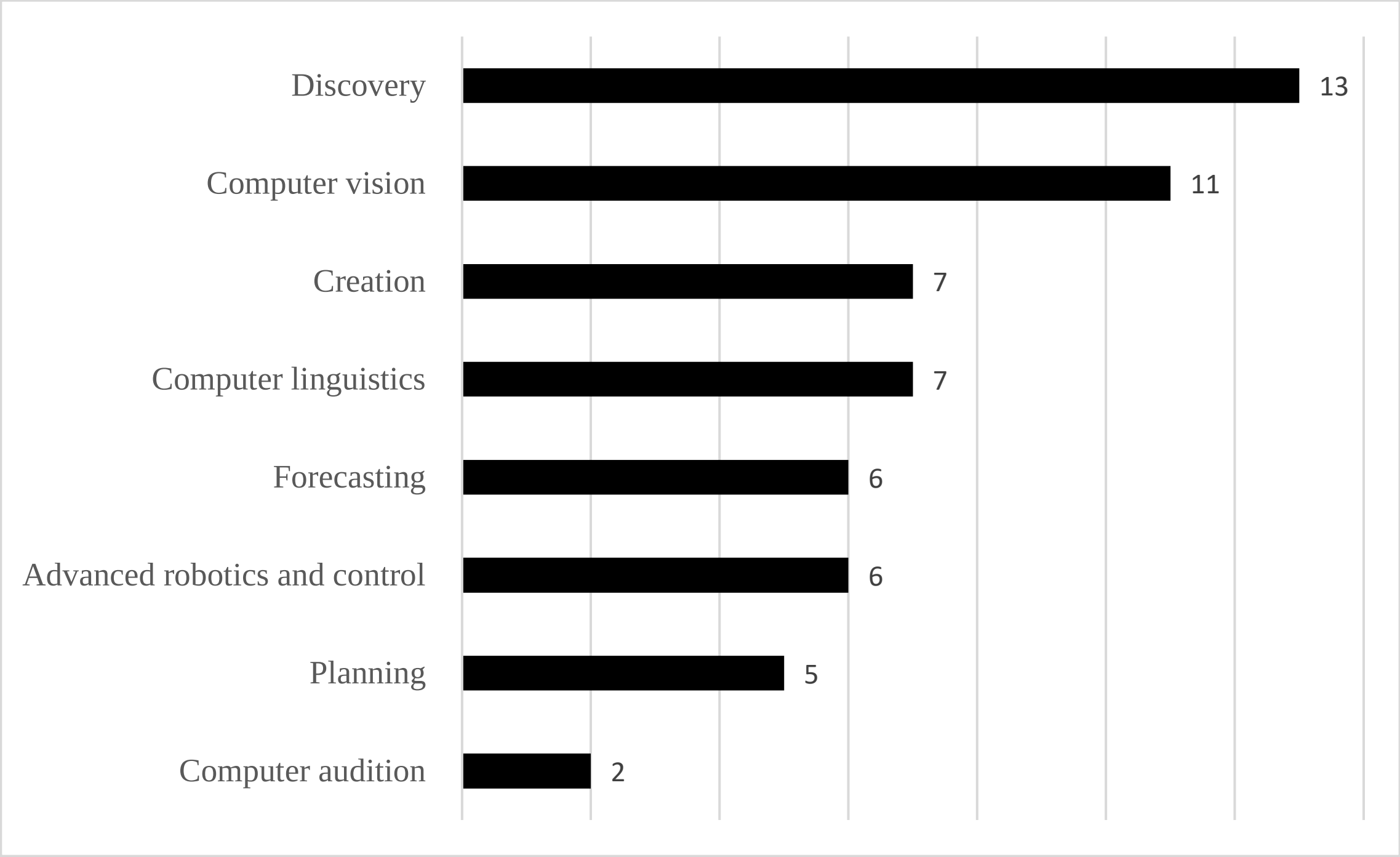

**FIGURE 3.** Overview of the capabilities of the 24 included AI-enabled systems

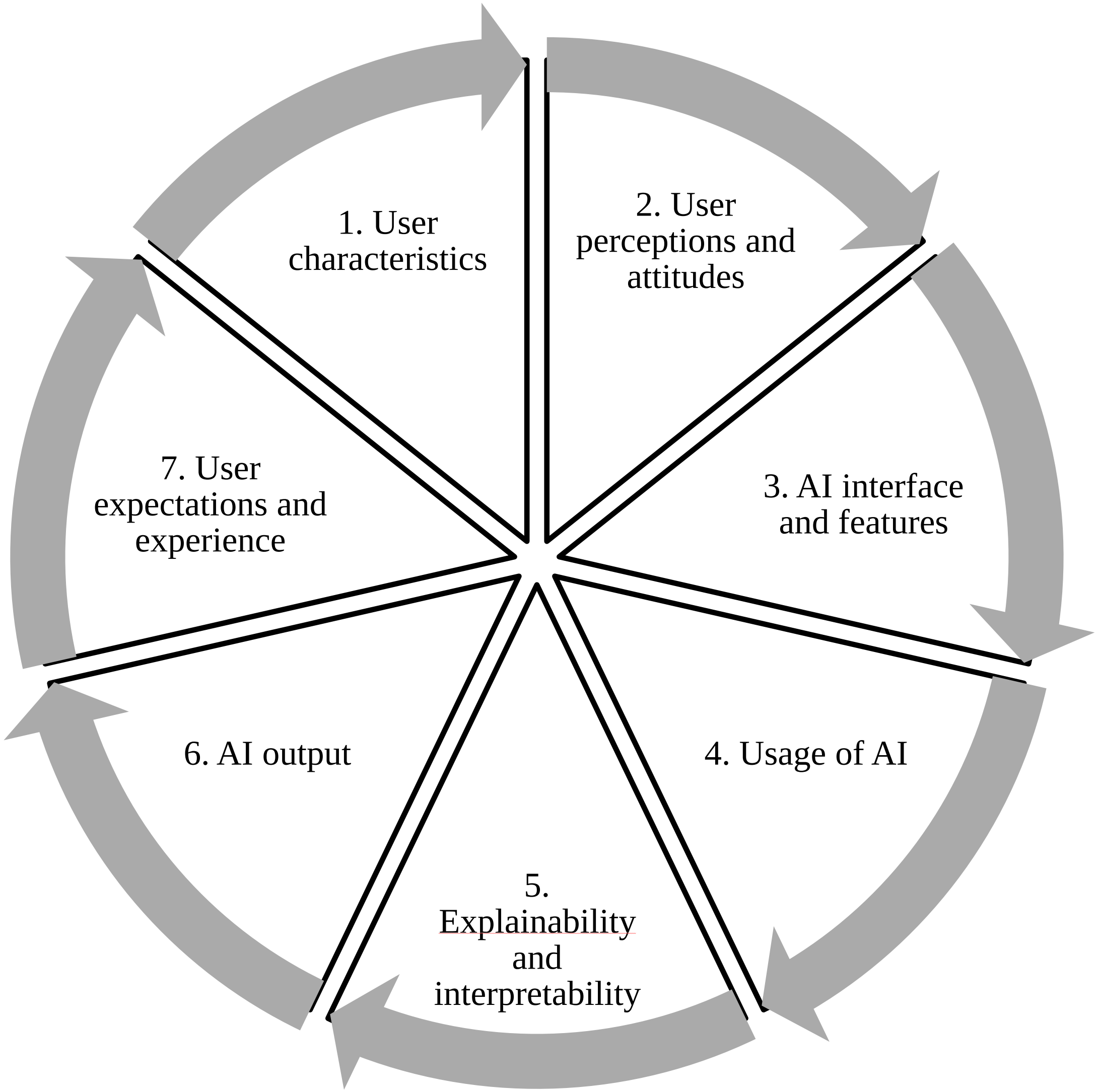

**FIGURE 4.** **Factors that influence human-AI interaction.**

**List of Tables**

TABLE I
The Inclusion and Exclusion Criteria

| **Inclusion criteria** | **Description** | **Justification** |
|---|---|---|
| Time of publication | January 1, 2011, to April 20, 2023 | We aimed to include relatively recent studies in this rapidly developing field |
| Language | English | Our search terms were English language words |
| Type | Peer-reviewed scientific articles or conference proceedings | Peer review to meet high level of evidence |
| Study methodology | Empirical | We aimed to collect evidence in the field |
| Study focus | Focused on the interaction between users and an AI-enabled system | Following the main goal of the review |
| Industry | Applied to a safety-critical industry | We were concerned with this high-risk field |
| User involvement | Involved users who intended to use the AI-enabled system in their study | A more general user profile would likely influence HAII in a different way than we intended |

TABLE II
The Search Terms and Justification

| **Search Terms** | **Justification based on the inclusion criteria** |
|---|---|
| 1. “optimise” OR “optimize” OR “improve” OR “advance” OR “effective”<br>AND | 1. To filter articles that focus on actionable efforts to foster more conducive HAII |
| 2. “artificial intelligence” OR “AI”<br>AND | 2. To filter articles that use AI-enabled systems |
| 3. “human” OR “humans” OR “user” OR “operator” OR “human factors”<br>AND | 3. To filter articles that focus on human-AI and involve humans or users |
| 4. “finding” OR “result”<br>AND | 4. To filter articles that are empirical |
| 5. “method” OR “tool” OR “technique” OR “practice”<br>AND | 5. To filter articles that are empirical |
| 6. “interaction” OR “interactions” OR “collaboration” OR “cooperation” OR “teaming” OR “teamwork” OR “integration”<br>AND | 6. To filter articles that focus on interaction between humans and AI |
| 7. “safety” OR “safe” OR “trust” | 7. To filter articles that focus on human-AI and applied in safety-critical industries or have a safety focus |

TABLE III
Summary of Main Findings

| Research questions | Summary of main findings |
| --- | --- |
| RQ1: What terms are used to describe the interaction between humans and an AI-enabled system? | • The term "interaction" covers a wide range of meanings.<br>• Several different terms are used, but it is unclear whether they have the same meanings.<br>• Seven of the 24 articles do not use any term.<br>• Future work should focus on whether and why terminologies matter and which ones best describe the interactions. |
| RQ2: What is the primary role of the AI-enabled system in the HAII? | • Based on the decision-making authority of AI-enabled systems (see Table VI), there are four primary roles: Collaborative AI, AI assisted decision-making, human controlled-AI, and emotional AI<br>• AI assisted decision-making, in which although users hold the final decision-making but AI-enabled systems can influence the process, is the most dominant primary role.<br>• Collaborative AI is the most equal decision-making authority between users and AI-enabled systems.<br>• Future work needs to explore more of the taxonomy of the primary roles to complete the list. |
| RQ3: What factors influence HAII? | • Seven factors influence HAII:<br>1. *User characteristics:* profiles and backgrounds of users that are found to affect HAII.<br>2. *User perceptions and attitudes*: perceptions and attitudes surrounding adopting AI-enabled systems into daily life, user biases, and consequences.<br>3. *User expectations and experience*: Gaps between user expectations and experience, and between AI and user conclusions.<br>4. *AI interface and features:* the importance of personalizing the design features and user interface of the AI-enabled system according to the users. For example, how an AI-enabled system communicates and interacts with users.<br>5. *Usage of AI:* how an AI-enabled system was used.<br>6. *Explainability and interpretability:* which AI rationale information should be communicated in what manner so that users accurately understand what is being communicated.<br>7. *AI output:* the information that should be presented to users, in which format and timepoints, and in what manner.<br>• Future work should focus on exploring whether and how HAII can be developed into human-AI relationships in safety-critical settings. Importantly, more research is needed to discover how the seven influencing factors found in this review can be used to improve HAII in safety-critical settings. |
| RQ4: How is HAII being measured in safety-critical industries? | • Quantitative and/or qualitative user-related outcomes are used as measurement variables.<br>• Most of the articles use subjective measurements of users (i.e., data collected from human responses to questionnaires or human observations and interviews).<br>• Objective measurements (i.e., data collected from physiological and physio-psychological measures and outcome measures) are used much less, and usually in combination with subjective measurements.<br>• Future work should focus on how to benefit from both subjective and objective measures to understand HAII better. |

TABLE IV
Overview of the 24 Included AI-Enabled Systems

| Article title (alphabetical order) | Description of the AI-enabled system (specific name if available) | Capabilities of the AI-enabled system [70] | The maturity of the AI-enabled system [71] |
|---|---|---|---|
| "Hello AI": Uncovering the Onboarding Needs of Medical Practitioners for Human-AI Collaborative Decision-Making [99] | System without an official product name: a Deep Neural Network (DNN) AI-enabled system as a support decision system for prostate cancer pathology | Computer Vision, Discovery | Proof-of-concept [TRL 3] |
| A Negotiation-Theoretic Framework for Control Authority Transfer in Mixed-Initiative Robotic Systems [91] | Expert-guided Mixed-Initiative Control Switcher (EMICS)<br><br>Negotiation-enabled Mixed-Initiative Control Switcher (NEMICS) | Advanced robotics and control, Planning | Prototype [TRL 4] |
| Algorithmic transparency and interpretability measures improve radiologists' performance in BI-RADS 4 classification [76] | System without an official product name: an AI-enabled system based on ResNet-50, a well-known convolutional neural network architecture, fine-tuned on mammograms from CBIS-DDSM dataset to classify lesions | Computer vision, Discovery | Proof-of-concept [TRL 3] |
| Can a chatbot enhance hazard awareness in the construction industry? [84] | Telegram: a chatbot | Computer linguistics; computer vision; creation | Commercial product/service (certified) [TRL 8] |
| Challenges in Designing a Fully Autonomous Socially Assistive Robot for People with Parkinson's Disease [83] | A social robot (NAO - an autonomous, programmable humanoid robot) | Advanced Robotics and Control, Computer Vision, Computer audition | Prototype [TRL 6] |
| Clinicians' Perceptions of an Artificial Intelligence–Based Blood Utilization Calculator: Qualitative Exploratory Study [94] | Blood Utilization Calculator (BUC), created by Integrated Vital Medical Dynamics LLC, is a proprietary AI-based tool designed to optimize the efficiency of blood transfusion practices. | Planning, Forecasting | Commercial product/service (certified) [TRL 8] |
| Co-design of human-centered, explainable AI for clinical decision support [86] | Doctor XAI enhances Doctor AI (a temporal model using recurrent neural networks (RNN) that predicts the patient's next visit time, diagnoses, and medications order from time stamped data) by integrating human-centered Explainable AI (XAI) technique | Discovery, Computer Linguistics, Creation | Prototype [TRL 4] |
| Designing Human-centered AI for Mental Health: Developing Clinically Relevant Applications for Online CBT Treatment [95] | System without an official product name: a clinical decision support tool that uses recurrent neural networks (RNNs) to offer personalized treatment insights for mental health conditions like depression and anxiety | Forecasting, Discovery | Prototype [TRL 4] |
| Designing Theory-Driven User-Centric Explainable AI [96] | System without an official product name: a clinical decision support tool | Forecasting, Discovery | Prototype [TRL 4] |
| Doctor's Dilemma: Evaluating an Explainable Subtractive Spatial Lightweight Convolutional Neural Network for Brain Tumor Diagnosis [78] | System without an official product name: a clinical decision support tool that leverages a novel Convolutional Neural Network integrated with Class Activation Mapping, designed for efficient and interpretable diagnosis of brain tumors from MRI images | Computer Vision, Discovery | Prototype [TRL 4] |
| Improving Workflow Integration with xPath: Design and Evaluation of a Human-AI Diagnosis System in Pathology [79] | xPath: a comprehensive and explainable human-AI collaborative diagnosis tool that can assist pathologists' examinations integrated into their practice | Computer Vision, Discovery | Prototype [TRL 6] |
| Indirect Shared Control Through Non-Zero Sum Differential Game for Cooperative Automated Driving [92] | System without an official product name: two neural networks — value NN for evaluation and Policy NN for control — are utilized within a dual iterative Approximate Dynamic Programming framework to optimize cooperation between human drivers and automated driving systems | Advanced Robotics and Control, Planning | Proof-of-concept [TRL 3] |
| Integrating machine learning predictions for perioperative risk management: | System without an official product name: ML-augmented risk assessment tool that utilizes SHAP values to deliver | Forecasting, Discovery | Prototype [TRL 4] |
| Towards an empirical design of a flexible-standardized risk assessment tool [93] | interpretable insights on the influence of individual features on its predictive outcomes. | | |
| Interactions between healthcare robots and older people in Japan: A qualitative descriptive analysis study [82] | Pepper: a social robot | Computer vision; computer audition; advanced robotics and control; computer linguistics; creation | Commercial product/service (certified) [TRL 8] |
| Lessons Learned from Designing an AI-Enabled Diagnosis Tool for Pathologists [80] | Impetus leverages Convolutional Neural Networks for feature extraction, Isolation Forest for outlier detection, and DBSCAN clustering to pinpoint areas of interest, streamlining tumor detection for pathologists on histological slides | Computer Vision, Discovery | Prototype [TRL 4] |
| PathNarratives: Data annotation for pathological human-AI collaborative diagnosis [85] | System without an official product name: AI-model for classification and captioning tasks for pathological human-AI collaborative diagnosis. | Computer vision; computer linguistics; creation | Prototype [TRL 4] |
| Psychophysiological Modeling of Trust In Technology: Influence of Feature Selection Methods [97] | System without an official product name: an autonomous vehicle was developed as a use case for testing Ensemble Trust Classifier Model, which utilizes EEG, EDA, and facial EMG data through MLP, SVM, and Gaussian Naïve Bayes algorithms for real-time trust prediction | Discovery, Forecasting | Proof-of-concept [TRL 3] |
| Rapid Assisted Visual Search: Supporting Digital Pathologists with Imperfect AI [81] | Rapid Assisted Visual Search (RAVS) employs convolutional neural networks (CNN) to facilitate medical assessments | Computer Vision, Discovery | Prototype [TRL 4] |
| Research on the influencing factors of user trust based on artificial intelligence self diagnosis system [87] | DoctorBot: an intelligent self-diagnosis system | Computer linguistics; Discovery, Creation | Commercial product/service (certified) [TRL 8] |
| Toward AI-enabled augmented reality to enhance the safety of highway work zones: Feasibility, requirements, and challenges [77] | System without an official product name: AI-enabled system that uses a fine-tuned YoloV4 neural network architecture for real-time vehicle detection/classification from distance in highways | Computer vision | Prototype [TRL 6] |
| Towards evaluating the impact of swarm robotic control strategy on operators' cognitive load [90] | Crazyflies- indoormicro-UAVs | Advanced robotics and control, Planning | Prototype [TRL 5] |
| Trust and Perceived Control in Burnout Support Chatbots [88] | System without an official product name: a burnout therapy chatbot, developed using Rasa, a Conversational AI platform for efficient AI assistant development | Computer Linguistics, Creation | Prototype [TRL 4] |
| User trust and understanding of explainable AI: Exploring algorithm visualisations and user biases [98] | Systems without an official product name: AI-enabled system with 3 ML algorithms (Decision Tree- DT, Logistic Regression - LR and Neural Networks - NN) for classification (Benign vs. Malignant) of biopsy results | Discovery | Proof of concept [TRL 3] |
| Why Do I Have to Take over Control? Evaluating Safe Handovers with Advance Notice and Explanations in HAD [89] | System without an official product name: AI-enabled system that uses description logic, automated planning, and natural language generation for predicting, explaining, and facilitating safe handovers from AI to human control in highly automated driving | Advanced Robotics and Control, Planning, Computer Linguistics, Creation, Forecasting | Prototype [TRL 4] |

TABLE V
Overview of the Factors Influencing Human-AI Interaction

| **Article title (alphabetical order)** | **Similar terms used for "interaction"** | **The primary role of the AI-enabled system in the interaction*** | **Factors influencing human-AI interaction** | | | | | | |
|---|---|---|---|---|---|---|---|---|---|
| | | | AI interface and features (N=13) | AI output (N=13) | Usage of AI (N=6) | Explainability and interpretability (N=12) | User expectations and experience (N=14) | User characteristics (N=10) | User perceptions and attitudes (N=16) |
| "Hello AI": Uncovering the Onboarding Needs of Medical Practitioners for Human-AI Collaborative Decision-Making [99] | "Collaboration" | AI assisted decision-making | | √ | | √ | √ | | √ |
| A Negotiation-Theoretic Framework for Control Authority Transfer in Mixed-Initiative Robotic Systems [91] | "Hand-offs" | Collaborative AI | √ | | | √ | | | |
| Algorithmic transparency and interpretability measures improve radiologists' performance in BI-RADS 4 classification [76] | "Interaction" and "Collaboration" | AI assisted decision-making | √ | √ | | | √ | √ | √ |
| Can a chatbot enhance hazard awareness in the construction industry? [84] | No term similar to "interaction" used | Human controlled-AI | | | √ | | | √ | |
| Challenges in Designing a Fully Autonomous Socially Assistive Robot for People with Parkinson's Disease [83] | "Assisting" | Emotional AI | | | | | √ | | √ |
| Clinicians' Perceptions of an Artificial Intelligence–Based Blood Utilization Calculator: Qualitative Exploratory Study [94] | No term similar to "interaction" used | AI assisted decision-making | | √ | √ | | √ | √ | |
| Co-design of human-centered, explainable AI for clinical decision support [86] | "Interaction" and "Collaboration" | AI assisted decision-making | √ | √ | | √ | √ | | √ |
| Designing Human-centered AI for Mental Health: Developing Clinically Relevant Applications for Online CBT Treatment [95] | "Partnership" | AI assisted decision-making | √ | √ | | √ | √ | | √ |
| Designing Theory-Driven User-Centric Explainable AI [96] | No term similar to "interaction" used | AI assisted decision-making | | | | √ | √ | | √ |
| Doctor's Dilemma: Evaluating an Explainable Subtractive Spatial Lightweight Convolutional Neural Network for Brain Tumor Diagnosis [78] | "Interaction" | AI assisted decision-making | √ | √ | | √ | √ | | √ |
| Improving Workflow Integration with xPath: Design and Evaluation of a Human-AI Diagnosis System in Pathology [79] | "Interaction" and "Collaboration" | AI assisted decision-making | √ | √ | | √ | √ | √ | √ |
| Indirect Shared Control Through Non-Zero Sum Differential Game for Cooperative Automated Driving [92] | "Interaction" and "Collaboration" | Collaborative AI | | | | | √ | | √ |
| Integrating machine learning predictions for perioperative risk management: Towards an empirical design of a flexible-standardized risk assessment tool [93] | No term similar to "interaction" used | AI assisted decision-making | √ | √ | √ | √ | | √ | |
| Interactions between healthcare robots and older people in Japan: A qualitative descriptive analysis study [82] | "Interaction" | Emotional AI | √ | | | | | √ | |
| Lessons Learned from Designing an AI-Enabled Diagnosis Tool for Pathologists [80] | "Interaction" and "Collaboration" | AI assisted decision-making | √ | √ | √ | √ | √ | | |
| PathNarratives: Data annotation for pathological human-AI collaborative diagnosis [85] | "Collaboration" | AI assisted decision-making | | | | √ | | | |
| Psychophysiological Modeling of Trust in Technology: Influence of Feature Selection Methods [97] | "Interaction" and "Collaboration" | Collaborative AI | | | √ | | √ | | √ |
| Rapid Assisted Visual Search: Supporting Digital Pathologists with Imperfect AI [81] | "Interaction" and "Collaboration" | AI assisted decision-making | √ | √ | √ | | √ | √ | √ |
| Research on the influencing factors of user trust based on artificial intelligence self diagnosis system [87] | No term similar to "interaction" used | AI assisted decision-making | √ | √ | | √ | | √ | √ |
| Toward AI-enabled augmented reality to enhance the safety of highway work zones: Feasibility, requirements, and challenges [77] | No term similar to "interaction" used | AI assisted decision-making | | | | | | √ | √ |
| Towards evaluating the impact of swarm robotic control strategy on operators' cognitive load [90] | "Interaction" | Human controlled-AI | | | | | | √ | √ |
| Trust and Perceived Control in Burnout Support Chatbots [88] | "Interaction" | AI assisted decision-making | √ | | | | √ | | √ |
| User trust and understanding of explainable AI: Exploring algorithm visualisations and user biases [98] | No term similar to "interaction" used | AI assisted decision-making | √ | √ | | √ | | | √ |
| Why Do I Have to Take over Control? Evaluating Safe Handovers with Advance Notice and Explanations in HAD [89] | "Handovers" | Collaborative AI | | √ | | | | | |

TABLE VI
Descriptions of the Primary Roles of AI-Enabled Systems in Human-AI Interaction

| Type of human-AI interaction | Description |
|---|---|
| Collaborative AI (N=4) | AI and users collaborating and cooperating to achieve a goal. |
| AI assisted decision-making (N=16) | Uses AI to assist users in making decisions by, for example, providing analyses, recommendations, or predictions. |
| Human controlled-AI (N=2) | Views AI merely as a tool to help humans to achieve a goal that requires human oversight, supervision, and intervention to function. |
| Emotional AI (N=2) | AI that can respond to user emotions. |

TABLE VII
Overview of the Participants Demographic Information, Dependent Variables, and Measurement Tools and Methods

| Article title (alphabetical order) | Industry | Study methodology | End-users (participants) | Number of participants (% female) | Mean of participant age | What was measured? | Measurement tools and methods |
|---|---|---|---|---|---|---|---|
| "Hello AI": Uncovering the Onboarding Needs of Medical Practitioners for Human-AI Collaborative Decision-Making [99] | Healthcare | A qualitative method | Pathologists | 21 (% female not stated) | Not stated | User perception of the AI-enabled system (what type of information users desired) | Qualitative: interviews |
| A Negotiation-Theoretic Framework for Control Authority Transfer in Mixed-Initiative Robotic Systems [91] | Nuclear energy | An experimental method | Robot operators | 10 (10% female) | 31.5 years old | • Performance: humans' task completion time*<br>• Number of collisions*<br>• Performance: number of human-AI conflicts*<br>• User perceptions of workload level | Quantitative:<br>• NASA TLX - quantitative interval scale [125]<br>• A free form qualitative usability questionnaire considering user acceptance, intuitiveness, and transparency of interaction |
| Algorithmic transparency and interpretability measures improve radiologists' performance in BI-RADS 4 classification [76] | Healthcare | A retrospective observer method | Radiologists | 4 (% female not stated) | Not stated | • The effect of different types of AI-based assistance on the radiologists' performance (sensitivity, specificity)<br>• User evaluation of the algorithm's assistance<br>• User attitudes towards AI integration into clinical workflow | • Qualitative: Interviews<br>• Quantitative: a combination of open-ended and multiple-choice questions in a questionnaire |
| | | | | | | Influence of personality traits | Quantitative: a publicly available personality test based upon the Big-Five personality model [126] and items from the International Personality Item Pool [127] |
| Can a chatbot enhance hazard awareness in the | Construction | An experimental method | Construction workers | 38 (% female not stated) | Not stated | User situational awareness* | • A physiological measure: eye movement |

| Article title (alphabetical order) | Industry | Study methodology | End-users (participants) | Number of participants (% female) | Mean of participant age | What was measured? | Measurement tools and methods |
|---|---|---|---|---|---|---|---|
| construction industry? [84] | | | | | | | • Qualitative: building good rapport<br>• Quantitative: a questionnaire for recruitment and gathering background information |
| Challenges in Designing a Fully Autonomous Socially Assistive Robot for People with Parkinson's Disease [83] | Healthcare | Experimental, qualitative, and quantitative methods | Patients with Parkinson's Disease | 10 (% female not stated) | Not stated | User perceptions of the AI-enabled system robot (e.g., robot functionality, user trust, physical and emotional support, social context) | Quantitative: a 19 item Likert-scale questionnaire<br><br>Qualitative: interviews |
| Clinicians' Perceptions of an Artificial Intelligence–Based Blood Utilization Calculator: Qualitative Exploratory Study [94] | Healthcare | A qualitative exploratory method | Clinicians | 10 (% female not stated) | Not stated | User perceptions of and trust in the AI-enabled system | Qualitative: interviews |
| Co-design of human-centered, explainable AI for clinical decision support [86] | Healthcare | An experimental method | Healthcare personnel | 41 participants (75.6% female) | 39 years old (SD=12) | • Implicit trust and confidence<br>• Explicit trust<br>• Behavioral intention and correlated constructs<br>• Explanation satisfaction<br>• User familiarity and involvement in the task<br>• Demographic information<br>• Need for cognition | • One 5-point Likert scale on user perceived AI reliability, predictability, and efficiency<br>• Unified theory of acceptance and use of technology (UTAUT) questionnaire [128]<br>• The Technology Acceptance Model (TAM) questionnaire [129]<br>• Explanation satisfaction scale [130]<br>• Familiarity and involvement in task scale [131]<br>• Need for Cognition scale [132], [133] |
| Designing Human-centered AI for Mental Health: Developing Clinically Relevant Applications for Online CBT Treatment [95] | Healthcare | A qualitative method | Healthcare personnel (Psychological Wellbeing Practitioners) | 13 (84% female) | Not stated | User evaluation of an AI-enabled system | Qualitative: interviews |
| Designing Theory-Driven User-Centric Explainable AI [96] | Healthcare | An experimental method | Clinicians | 14 clinicians (71,4% female) | 24-29 years old | • Demographic information<br>• How users interacted and used the AI-enabled system (user reasoning process and perceptions) | • A brief background survey |

| Article title (alphabetical order) | Industry | Study methodology | End-users (participants) | Number of participants (% female) | Mean of participant age | What was measured? | Measurement tools and methods |
|---|---|---|---|---|---|---|---|
| Doctor's Dilemma: Evaluating an Explainable Subtractive Spatial Lightweight Convolutional Neural Network for Brain Tumor Diagnosis [78] | Healthcare | Experimental and quantitative methods, and usability testing | Clinicians | 10 (% female not stated) | Not stated | • Task completion time*<br>• Task completion rate<br>• AI-enabled system's accuracy*<br>• Task difficulty feedback<br>• User perceptions of the AI-enabled system | Quantitative: Expert Survey Evaluation |
| Improving Workflow Integration with xPath: Design and Evaluation of a Human-AI Diagnosis System in Pathology [79] | Healthcare | Experimental, qualitative, and quantitative methods | Clinicians | 12 (% female not stated) | Not stated | • AI-enabled system's accuracy*<br>• Efficiency: Time spent using the AI-enabled system vs. a traditional interface*<br>• Perceived effort and workload<br>• User perceptions and attitude towards the AI-enabled system | Quantitative:<br>• The College of American Pathologists (CAP) cancer protocol template (to diagnose) [134]<br>• A post-questionnaire with a Likert scale |
| Indirect Shared Control Through Non-Zero Sum Differential Game for Cooperative Automated Driving [92] | Automotive | Experimental and quantitative methods | Drivers | 11 (% female not stated) | Not stated | • User lateral errors and steering wheel angular acceleration*<br>• Perceived workload<br>• User subjective feeling | Quantitative:<br>• NASA-Task Load Index (NASA-LTX) [135], [125]<br>• The six-dimension 5-point Likert scale |
| Integrating machine learning predictions for perioperative risk management: Towards an empirical design of a flexible-standardized risk assessment tool [93] | Healthcare | A qualitative method | Healthcare personnel | 17 (35.29% female) | Not stated | • Level of agreement between ML's and users' risk ranking<br>• User perceptions of design preferences<br>• User perceptions of the potential impact of the ML tool on the clinical workflow | Qualitative:<br>• Cognitive walkthrough<br>• Interviews |
| Interactions between healthcare robots and older people in Japan: A qualitative descriptive analysis study [82] | Healthcare | A qualitative descriptive analysis study | Healthcare personnel and patients | 23 (% female not stated) | Not stated | Interaction between the robot and users (healthcare personnel and patients) | Qualitative: observations |
| Lessons Learned from Designing an AI-Enabled Diagnosis Tool for Pathologists | Healthcare | Experimental and qualitative methods | Pathologists | 8 (% female not stated) | Not stated | User evaluation of the AI-enabled system | Qualitative: interviews |
| PathNarratives: Data annotation for pathological human-AI | Healthcare | An experimental method | Pathologists | 9 (% female not stated) | Not stated | User trust in the algorithm for classification | Quantitative: a Likert scale questionnaire |

| Article title (alphabetical order) | Industry | Study methodology | End-users (participants) | Number of participants (% female) | Mean of participant age | What was measured? | Measurement tools and methods |
|---|---|---|---|---|---|---|---|
| collaborative diagnosis [85] | | | | | | | |
| Psychophysiological Modeling of Trust in Technology: Influence of Feature Selection Methods [97] | Automotive | Experimental and quantitative methods | Drivers | 25 (35% female) | 27.55 years old | • User trust in the AI-enabled system<br>• Logs comprised the trust-related behavior (Number of time AI vs. manual control was used)* | Quantitative: Trust in Technology questionnaire [136], [137], [138] |
| Rapid Assisted Visual Search: Supporting Digital Pathologists with Imperfect AI [81] | Healthcare | Experimental and qualitative methods | Pathologists | 6 (% female not stated) | Not stated | • Efficiency: time taken per section, derived from usage logs*<br>• Users' positive/negative counting and scoring sections<br>• User trust and evaluation of the AI-enabled system | Qualitative: interviews |
| Research on the influencing factors of user trust based on artificial intelligence self-diagnosis system [87] | Healthcare | Experimental and quantitative methods | Patients | Study 1: 13 (23.07%)<br><br>Study 2: 48 (% female not stated)<br><br>Total: 61 | Study 1: 40.46 years<br><br>Study 2: Not stated | User perceptions of and trust in the AI-enabled system | Quantitative: a 10-point Likert-scale questionnaire |
| Toward AI-enabled augmented reality to enhance the safety of highway work zones: Feasibility, requirements, and challenges [77] | Construction | Quantitative and qualitative methods | Highway construction workers (maintenance crew and affiliated members) | 129 (% female not stated) | Not stated | User perceptions of the AI-enabled system | • Qualitative: an initial interview, cognitive walkthrough<br>• Quantitative: a questionnaire |
| Towards evaluating the impact of swarm robotic control strategy on operators' cognitive load [90] | Space | An experimental method | Drone operators | 40 (32.50% female) | Not stated | User cognitive load* | Psychophysiological methods: Biopac MP35 with a 3-lead ECG electrode set |
| | | | | | | User perceptions of workload | Quantitative: a modified NASA TLX [139] |
| Trust and Perceived Control in Burnout Support Chatbots [88] | Healthcare | Experimental and quantitative methods | Patients | 35 (63,33% female) | Not stated | • User trust<br>• Perceived control<br>• Usability<br>• The unconscious attribution of human characteristics to artificial agents<br>• Perceived closeness<br>• User attitude | Quantitative:<br>• The Human-Computer Trust scale [138]<br>• The Yu scale (perceived control) [140]<br>• UMUX-LITE scale [141]<br>• Humanness or Mindless Anthropomorphism scale [142] |

| Article title (alphabetical order) | Industry | Study methodology | End-users (participants) | Number of participants (% female) | Mean of participant age | What was measured? | Measurement tools and methods |
|---|---|---|---|---|---|---|---|
| | | | | | | | • The Inclusion of the Other in the Self (IOS) scale [143]<br>• Attitude Towards AI: Two, five-point Likert scale items borrowed from 20-item questionnaire proposed by [144] |
| User trust and understanding of explainable ai: Exploring algorithm visualisations and user biases [98] | Healthcare | An experimental method | Participants some with and some without experience working in healthcare | 70 (62.90% female) | 30.06 years | User perceptions of and trust in the AI decision support system | Quantitative: an online Qualtrics survey |
| Why Do I Have to Take over Control? Evaluating Safe Handovers with Advance Notice and Explanations in HAD [89] | Automotive | Experimental, quantitative, and qualitative methods | Drivers | 23 (65.21% female) | 25.13 years | User trust in automation | Quantitative: Trust in Automation Questionnaire (TiQ) [145] |
| | | | | | | User perceptions of workload | Quantitative: NASA Task Load Index (NASA TLX) [135] |
| | | | | | | User handover condition preferences | Qualitative: interviews |

*Objective measures

## Appendix 1. The Search Strings of All Four Digital Databases

### Scopus

(TITLE-ABS-KEY(Optimise OR optimize OR improve OR advance OR effective)) AND (TITLE-ABS-KEY(AI OR "Artificial intelligence")) AND (TITLE-ABS-KEY(human OR user OR operator OR humans OR "human factors")) AND (TITLE-ABS-KEY(finding OR result)) AND (TITLE-ABS-KEY(Method OR tool OR technique OR practice)) AND (TITLE-ABS-KEY(interaction OR collaboration OR cooperation OR teaming OR teamwork OR interactions OR integration)) AND (TITLE-ABS-KEY(safety OR safe OR trust)) AND ( LIMIT-TO ( PUBYEAR,2023) OR LIMIT-TO ( PUBYEAR,2022) OR LIMIT-TO ( PUBYEAR,2021) OR LIMIT-TO ( PUBYEAR,2020) OR LIMIT-TO ( PUBYEAR,2019) OR LIMIT-TO ( PUBYEAR,2018) OR LIMIT-TO ( PUBYEAR,2017) OR LIMIT-TO ( PUBYEAR,2016) OR LIMIT-TO ( PUBYEAR,2015) OR LIMIT-TO ( PUBYEAR,2014) OR LIMIT-TO ( PUBYEAR,2013) OR LIMIT-TO ( PUBYEAR,2012) OR LIMIT-TO ( PUBYEAR,2011) )

### Web of Science

ALL=((“optimise” OR “optimize” OR “improve” OR “advance” OR “effective”) AND (“artificial intelligence” OR “AI”) AND (“human” OR “humans” OR “user” OR “operator” OR “human factors”) AND (“finding” OR “result”) AND (“method” OR “tool” OR “technique” OR “practice”) AND (“interaction” OR “interactions” OR “collaboration” OR “cooperation” OR “teaming” OR “teamwork” OR “integration”) AND (“safety” OR “safe” OR “trust”))

### IEEE Xplorer

("All Metadata":Optimise OR "All Metadata":optimize OR "All Metadata":improve OR "All Metadata":advance OR "All Metadata":effective) AND ("All Metadata":AI OR "All Metadata":"Artificial intelligence") AND ("All Metadata":human OR "All Metadata":user OR "All Metadata":operator OR "All Metadata":"human factors") AND ("All Metadata":finding OR "All Metadata":result) AND ("All Metadata":method OR "All Metadata":tool OR "All Metadata":technique OR "All Metadata":practice) AND ("All Metadata":interaction OR "All Metadata":collaboration OR "All Metadata":cooperation OR "All Metadata":teaming OR "All Metadata":teamwork OR "All Metadata":integration) AND ("All Metadata":safety OR "All Metadata":safe OR "All Metadata":trust)
Filters Applied: 2011-2023

### ACM digital library

[[Abstract: "artificial intelligence"] OR [Abstract: "ai"]] AND [[Abstract: "method"] OR [Abstract: "tool"] OR [Abstract: "technique"] OR [Abstract: "practice"]] AND [[Abstract: "human"] OR [Abstract: "humans"] OR [Abstract: "user"] OR [Abstract: "operator"] OR [Abstract: "human factors"]] AND [[Full Text: "finding"] OR [Full Text: "result"]] AND [[Full Text: "interaction"] OR [Full Text: "interactions"] OR [Full Text: "collaboration"] OR [Full Text: "cooperation"] OR [Full Text: "teaming"] OR [Full Text: "teamwork"] OR [Full Text: "integration"]] AND [[Full Text: "optimise"] OR [Full Text: "optimize"] OR [Full Text: "improve"] OR [Full Text: "advance"] OR [Full Text: "effective"]] AND [[Full Text: "safety"] OR [Full Text: "safe"] OR [Full Text: "trust"]] AND [E-Publication Date: (01/01/2011 TO 04/30/2023)]